\documentclass[10pt]{article}
\usepackage{amsmath,amssymb,bm,epsf,epsfig,graphicx}
\input epsf.sty
\usepackage[utf8]{inputenc}
\usepackage[english]{babel}
\usepackage{amsmath}
\usepackage{latexsym}
\usepackage{amsfonts}
\usepackage{mathtools}
\usepackage{amssymb}
\usepackage{scalerel}
\usepackage{extpfeil}
\usepackage[left=3cm,right=3cm,top=3.1cm,bottom=3.1cm]{geometry}
\usepackage{cite}
\usepackage{tensor}
\usepackage{tikz}
\usepackage{tikz-cd}
\usepackage{bbm}

\usepackage[mathscr]{euscript}
\usetikzlibrary{arrows.meta, bending}
\tikzset{C/.style={circle, minimum size=8mm,
		node contents={},
		append after command={\pgfextra{%
				\draw[-{Straight Barb[flex']}](\tikzlastnode.150) arc (450:110:2.8mm);}
	}}
}

\usepackage{hyperref}
\numberwithin{equation}{section}

    \newcommand{\Tr}{\mathop{\rm Tr}\nolimits}

    \def\p{\partial}

    \def \be {\begin{eqnarray}}
    \def \ee {\end{eqnarray}}
    \def \bal {\begin{align}}
    \def \eal {\end{align}}
    \def \bdm {\begin{displaymath}}
    \def \edm {\end{displaymath}}

    \def\0{\nonumber}

    \def\wc{\omega_\text{c}}
    \def\wo{\omega_\text{o}}
    
   \newcommand{\ch}[1]{{\color{black}{#1}}}

\begin{document}
	\begingroup\allowdisplaybreaks

\vspace*{1.1cm}

\centerline{\Large \bf Bulk-induced D-brane deformations }
\vspace {.3cm}

\centerline{\Large \bf and the string coupling constant} 

\vspace{.3cm}

\begin{center}

{\large Carlo Maccaferri$^{(a)}$\footnote{Email: maccafer at gmail.com}, Alberto Ruffino$^{(a)}$\footnote{Email: ruffinoalb at gmail.com}  and  Jakub Vo\v{s}mera$^{(b)}$\footnote{Email: jakub.vosmera at ipht.fr} }
\vskip 1 cm
$^{(a)}${\it Dipartimento di Fisica, Universit\`a di Torino, \\INFN  Sezione di Torino \\
Via Pietro Giuria 1, I-10125 Torino, Italy}
\vskip .5 cm
\vskip .5 cm
$^{(b)}${\it Institut de Physique Théorique\\
	CNRS, CEA, Université Paris-Saclay\\
 Orme des Merisiers, Gif-sur-Yvette, 91191 CEDEX, France}

%
\end{center}

\vspace*{6.0ex}

\centerline{\bf Abstract}
\bigskip
We consider \ch{computing} the on-shell disk action of open-closed string field theory as a gauge-invariant way of capturing the shift in D-brane tension that is induced by a deformation of the bulk CFT. We study the effect of bulk \ch{matter} deformations (both marginal and relevant) on a wide range of boundary conditions in a number of CFTs \ch{up to subleading (two-loop) order in perturbation theory}. In all analyzed examples, we find that the shift in the $g$-function \ch{of the matter boundary state} is always accompanied by a \ch{boundary-independent} shift in the string coupling constant, whose leading behaviour is universally proportional to the sphere two-point function of the deforming bulk operator.

\baselineskip=16pt
\newpage
\tableofcontents

\section{Introduction}\label{sec:1}
It is known since early times that string field theory (SFT)  provides a way to investigate the space of 2d (B)CFT's \cite{Mukherji:1991tb, Mukherji:1991kz, Freedman:2005wx, Kudrna:2012re} by connecting different conformal vacua  via classical solutions. This has recently received a renewed attention \cite{Scheinpflug:2023osi, Scheinpflug:2023lfn, Mazel:2024alu} as an alternative convenient approach to the standard conformal perturbation theory (CPT) \cite{Zamolodchikov:1986gt,Zamolodchikov:1987ti,Zamolodchikov:1989hfa,Affleck:1991tk,Friedan:2003yc}. 
The advantage of SFT versus CPT is that it provides an automatically consistent way of regulating  singularities from collisions of perturbing fields by recognizing such collisions as originating from degenerations of punctured Riemann surfaces (in this particular application spheres and disks). In this picture the collisions of the perturbing  operators happen  at a corner of moduli space that is always associated to a SFT Feynman diagram, with the SFT propagator taking care of the degenerating region in a way which is analogous to the $i\epsilon$ prescription of quantum field theory \cite{Sen:2019jpm, Larocca:2017pbo, Witten:2013pra}. In this way all contact divergences are systematically regulated.
Along this direction, following the  pioneering analysis of \cite{Mukherji:1991tb}, there has been progress \cite{Scheinpflug:2023lfn, Mazel:2024alu} in the study of particular examples of bulk RG flows triggered by a slightly relevant operator $\mathbb{V}$ of conformal dimensions $(1-y,1-y)$   (with $y$ very small and positive) whose OPE closes on itself
\begin{align}
\mathbb{V}(z,\bar z)\,\mathbb{V}(0,0)=\frac1{|z|^{4(1-y)}}+C_{\mathbb{VVV}}\,\frac {\mathbb{V}(0,0)}{|z|^{2(1-y)}}+ ({\rm regular})\,.
\end{align}
Calling $t$ the coefficient  (coupling) of the deforming field  $T=t(y) \mathbb{V}$, in this case the leading order beta function has the structure
\begin{align}
\beta_t\sim 2y\,t(y)-C_{\mathbb{VVV}} \,t(y)^2+\ldots\,,
\end{align}
and therefore we can search for a perturbative solution in the small parameter $y$ so that a new non-trivial conformal point ($\beta=0$) is found at
\begin{align}
t(y)\sim \frac{2y}{C_{\mathbb{VVV}}}+\mathcal{O}(y^2)\,.
\end{align}
This represents an IR fixed point which is parametrically close to the initial perturbative vacuum $t=0$ (UV fixed point) and we refer to this  as a {\it short} RG-flow.   The paradigm example of short RG flows are the Zamolodchikov minimal model flows triggered by the relevant field $\mathbb{V}=\phi_{(1,3)}$, which connect the $m$-th (UV) to the $(m-1)$-th  (IR) minimal models, in the large $m\sim2/y$ limit \cite{Zamolodchikov:1987ti}.

Going to higher order in $y$  in CPT is notoriously hard and the details of the calculation depend on the chosen regularization scheme, see for example \cite{Gaberdiel:2008fn}.  On the other hand, the analogous analysis in SFT requires to solve classical (from the target space perspective) equations of motion which are concretely defined, once the SFT is constructed. This is carried out by embedding the CFT of interest in a critical $c=26$ matter CFT, together with the $c=-26$ $bc$ system and searching for a solution 
$$
\Phi(y)\sim y t_1 c\bar c \mathbb{V}+\mathcal{O}(y^2)\,,
$$ 
solving order by order in $y$ the classical equation of motion.  The SFT solution $\Phi(y)$ is a state in the full Hilbert space of the combined matter-ghost  $c=26-26=0$ CFT and it should be thought of as a collection of vacuum expectation values for the states that are responsible for the RG flow plus an infinite dressing of irrelevant and auxiliary fields which are needed to solve the full SFT equation of motion. In particular, after having integrated out these extra fields, the equations of motion for the relevant fields should be thought of as the SFT incarnations of the $\beta$ functions of CPT, where the ambiguities related to the definition of the regularization scheme on the CPT side are reflected in the freedom of choosing the SFT data and gauge fixing. The conformal data of the IR fixed point are expected to be encoded in gauge invariant quantities involving the (non gauge invariant) classical solution $\Phi(y)$. For example the physical fluctuations around the solution are expected to encode the conformal weights of primaries at the IR fixed point. Another natural gauge invariant quantity is the classical SFT action. In closed string field theory however the value of the action on a classical solution is expected to formally vanish \cite{Erler:2022agw} because of the existence of the ghost-dilaton $D=c\partial^2 c-\bar c\bar\partial^2 \bar c$ which can be used to arbitrarily rescale the action \cite{Bergman:1994qq}. Nevertheless an explicit computation \cite{Scheinpflug:2023lfn} recently showed that the action evaluated on the $\Phi(y)$ classical solution appears to be universally proportional to the shift in the central charge
\begin{align}
S[\Phi(y)]=-\frac1{12}\Delta c=\mathcal{O}(y^3)\,.
\end{align}
This has been non-trivially verified up to $\mathcal{O}(y^4)$ by comparing with the known expectation from Zamolodchikov flows of minimal models.
This seems to be at odds with the expected vanishing action from \cite{Erler:2022agw} and indeed a closer inspection shows that $\Phi(y)$ is not really a solution in that it violates the equation of motion in the direction of the dual ghost dilaton state $\tilde D=c_0^+D$
\begin{align}
{\rm EOM}(\Phi(y))\sim y^3 \tilde D+\ldots\,. \label{gdobs}
\end{align}
This problem was realized already in \cite{Mukherji:1991tb} and attributed to the fact that the solution is trying to change the central charge of the mater CFT while string theory requires a critical fixed central charge. This basic tension is manifestated in the ghost dilaton obstruction \eqref{gdobs}.  However, as diccussed in \cite{Mukherji:1991tb, Mazel:2024alu}, this obstruction can be trivialized by assuming the existence of a linear dilaton factor in the matter CFT. If this factor is present, then $\tilde D$ can be trivialized by an allowed ghost number two state $\Theta$ (see \eqref{eq:theta})
\begin{align}
\tilde D=\frac1{\beta} Q_{\rm c}\Theta\,,
\end{align}
where $\beta$ is the background charge of the linear dilaton and $Q_{\rm c}$ the closed string BRST operator.
This then allows to correct the $\Phi(y)$ solution by adding a $y^3\Theta$ term which cancels the ghost-dilaton obstruction. Moreover $\Theta$ has the effect of shifting the linear dilaton background charge and thus the central charge so that the full solution now describes a flow of central charge from the initial perturbed CFT to the linear dilaton factor in such a way that the total central charge stays constant and the full string theory background remains critical. It remains to be seen what are the consequences of this on the closed SFT action, but this is not the direction we want to take in this paper.

In general, a closed SFT solution will not just change the bulk CFT background but also the string coupling constant $g_\mathrm{s}$. Detecting a change of the string coupling via a gauge invariant observable in closed string field theory does not  seem to be straightforward, though. In this regard it was pointed out in \cite{cosmo} that one can use probe D-branes for this task. This can be done by adding an open string sector to the closed SFT and studying how the open string vacuum state  $\Psi=0$ is shifted to a new solution $\Psi^*$ as a response to the change in the closed string background induced by a given closed string solution $\Phi^*$. The natural observable that is associated to this process is the combined disk vacuum energy generated by the closed SFT solution $\Phi^*$, together with the open string vacuum shift $\Psi^*$. This quantity $\Lambda(\Phi^*,\Psi^*)$ was conjectured in \cite{cosmo} to compute the shift in the string theory disk partition function 
\begin{align}
   \Lambda(\Phi^\ast,\Psi^\ast)=  -\frac{1}{2\pi^2} \bigg(\frac{g^\ast}{g_\mathrm{s}^\ast}-\frac{g}{g_\mathrm{s}}\bigg)\,, \label{gensen}
\end{align}
where $g$ and $g^\ast$ are the $g$-functions of the initial and final boundary states and $g_\mathrm{s}$ and $g_\mathrm{s}^\ast$ are the initial and final string coupling constants.  We call this the  {\it generalised}  Sen's conjecture. 
 By checking \eqref{gensen} with a sufficient number of boundary states subject to the same closed string field theory solution one can  isolate a universal part in $\Lambda(\Phi^\ast,\Psi^\ast)$ (independent of the boundary conditions) which represents the change in $g_s$. In this paper we perform this analysis in two rather different cases and observe that the change in the string coupling constant is universally the same, in the sense to be described below. 
 
 To start with, extending the analysis of \cite{cosmo}, we compute the on-shell disk action for various boundary states subject to a marginal deformation of a Narain lattice  of $d$ free bosons. By comparing with the expected answer of the deformed $g$-functions from the exact (B)CFT sigma model, we are able to detect that the bulk solution is changing the string coupling constant as described in \eqref{narain-gs}.
 
  In the second, and more important, application we study how a short bulk  RG flow affects the initial boundary state on which we define the open-closed SFT. \ch{Under the assumption that the induced boundary flow is also short,} we are able to find a closed-form expression for the gauge invariant quantity $\Lambda(\Phi^\ast,\Psi^\ast)$ that is exact up to $\mathcal{O}(y^2)$ (thus one order beyond the CPT analysis of \cite{Fredenhagen:2006dn}). This gives
  \begin{equation}
    \begin{split}
    &\frac{g^\ast}{g}\bigg(\frac{g_\mathrm{s}^\ast}{g_\mathrm{s}}\bigg)^{-1}=1+ \frac{\tensor{B}{_{\mathbb{V}\mathbbm{1}}}}{g}\left(\frac{y}{    \tensor{C}{_{\mathbb{VVV}}} 
  }\right)+
  \left(\frac{\tensor{B}{_{\mathbb{V}\mathbbm{1}}}}{g}\frac{2{\cal A}^{\rm f.p.}_{TTTT}}{3 \tensor{C}{_{\mathbb{VVV}}}}+\Tilde{{\cal A}}^{\rm f.p.}_{TT}
  \right)\left(\frac{y}{    \tensor{C}{_{\mathbb{VVV}}} 
  }\right)^2+\mathcal{O}(y^3)\,, \label{main1}
    \end{split}
\end{equation}
where ${\cal A}^{\rm f.p.}_{TTTT}$ is the regularized zero-momentum bulk four-point amplitude computed in \cite{Scheinpflug:2023lfn}, $\tensor{B}{_{\mathbb{V}\mathbbm{1}}}$ is the bulk-boundary OPE between the bulk deforming field and the boundary identity and finally $\Tilde{{\cal A}}^{\rm f.p.}_{TT}$ is the regularized disk 2-point amplitude defined in \eqref{atildett}. This is the first main result of this paper.

To extract the change in the string coupling constant triggered by the bulk relevant solution, we apply this formula to particular Cardy boundary states in large $m$ Virasoro minimal models \ch{$\text{MM}_m$} subject to the Zamolodchikov short RG flow triggered by \ch{the field} $\phi_{(1,3)}$ and compare with the known exact changes in the $g$-functions. This can be done independently for two different classes of boundary conditions and we unambiguosly find that the string coupling constant changes as \eqref{gs-minimal}.
  
   In both the marginal and relevant cases the closed string solution is perturbative in the sense that it admits a power series expansion, either in the marginal parameter (the Narain modulus) or in the small parameter $y$ associated to the small RG-flow $\text{MM}_m\to \text{MM}_{m-1}$. In both cases we can thus write
 \begin{align}
 \Phi^*=c\bar c T+({\rm subleading})\,,
 \end{align}
 where $T$ is a pure-matter marginal or (slightly) relevant field with normalization which is free (but small) for marginal deformations or fixed by the equation of motion to a (small) value, in the case of relevant deformations. In both cases when we compute the on-shell disk action for a set of different boundary conditions as described above we observe that the change in the string coupling constant is universally given by
 \begin{align}
 \frac {g_\mathrm{s}^\ast}{g_\mathrm{s}}=1+\frac18 \frac{\langle T,T\rangle}{\langle0|0\rangle}+({\rm subleading})\,,\label{main2}
 \end{align}
where $\langle \cdot,\cdot\rangle$ is the BPZ inner product in the matter bulk CFT and $|0\rangle$ is its $SL(2,{\mathbb C})$ vacuum. This is the other central observation of this paper. Notice that this is eventually a statement about the purely closed string sector without any reference to the open string sector which, from this perspective,  has only been used as an auxiliary ingredient.

The paper is organized as follows. In section \ref{sec:2} we review the main results of \cite{cosmo} about the \ch{properties} of the \ch{on-shell} disk action \ch{of} open-closed SFT. We then specialize to the case of bulk marginal deformations and, in particular, to deformations of a Narain lattice of $d$ free bosons. By computing the disk action \ch{of a solution describing a deformation of the Narain modulus in the presence of generic boundary states}, we are then able to identify how the string coupling constant is changed by the \ch{closed-string part} of the SFT solution. In section \ref{sec:3} we apply our framework to the case of short bulk RG-flows, consistently setting up a bulk-boundary perturbative scheme in order to associate an open-string vacuum-shift solution to a given closed SFT solution. We discuss under which conditions the open-string vacuum-shift solution is perturbative (as the seeding closed SFT solution is) and we write down the closed- and the open-string solutions to compute the disk action up to subleading order. \ch{This would} correspond to \ch{a two-loop calculation} in CPT. Section \ref{sec:4} contains the main technical results of this paper and it is fully devoted to the computation of the disk action for small RG flows. We \ch{perform} the computation in the limit of large closed-string stubs and, as a non-trivial consistency check of our construction, \ch{we show that} our disk action \eqref{main1} turns out to be fully independent of the SFT data used to define the interaction vertices of the open-closed SFT. In section \ref{sec:5}, we test our general result for the disk action against the expected change in the boundary entropy for a set of Cardy boundary states in large $m$ minimal models subject to Zamolodchikov short RG flows triggered by \ch{the field} $\phi_{(1,3)}$. After matching the $g$-function part of the disk action, we unambiguously identify a second order change in the string coupling constant, which \ch{we find to be fully parallel} to the change that was detected in the case of the marginally deformed Narain lattice. In the concluding section \ref{sec:6} we further discuss our main results and we outline various research directions suggested by our work. \ch{Finally}, in appendix \ref{app:A}, we compute (for the first time, to our knowledge) the bulk two-point function of \ch{the field} $\phi_{(1,3)}$ on a disk in the large $m$ limit by explicitly solving the third order differential equation implied by the existence of a level-three null state and by fixing the integration constants using the known bulk-boundary BCFT data.

\section{The on-shell disk action}\label{sec:2}

In this section we will review the main results and observations of \cite{cosmo} relating to the disk part of the open-closed SFT action evaluated on a classical closed string-field configuration $\Phi^\ast$, as well as on an open-string field configuration $\Psi^\ast$ which can be found as a classical solution of open-string equations of motion on the closed-string background given by $\Phi^\ast$.

\subsection{Generalised Sen's conjecture}

In the context of pure open-string field theory, it has been conjectured by Ashoke Sen that the on-shell action gives a direct measure of the difference between the $g$-functions (tensions) of the D-brane associated with the perturbative vacuum $\Psi=0$ and the D-brane which is represented by a classical open-string field configuration $\Psi^\ast$. We will now discuss a generalisation of this statement to the situations where one is allowed to also change the closed-string background by turning on a non-trivial classical closed-string field configuration $\Phi^\ast\neq 0$.

\subsubsection{Changing the closed-string background}\label{sec:2.1.1}

Let us start by writing down the pure-closed classical string field theory action \cite{Zwiebach:1992ie}
\begin{align}
    S_\text{sphere}[\Phi] = Z_\mathrm{sphere}-\frac{1}{g_\mathrm{s}^2}\sum_{k=1}^\infty \frac{1}{(k+1)!}\omega_\mathrm{c}\big(\Phi,l_k(\Phi^{\wedge k})\big)\,.\label{eq:CSFTAction}
\end{align}
Here, the degree-even dynamical string field $\Phi\in \mathcal{H}_\mathrm{c}$ is a linear combination of states in the combined matter and ghost CFT which determine the worldsheet theory of a closed string propagating on a particular (classical) background. Furthermore, $\omega_\mathrm{c}: \mathcal{H}_\mathrm{c}\otimes  \mathcal{H}_\mathrm{c}\longrightarrow \mathbb{C}$ denotes a symplectic form given by the BPZ product on $\mathcal{H}_\mathrm{c}$, while $l_k: (\mathcal{H}_\mathrm{c})^{\otimes k}\longrightarrow \mathcal{H}_\mathrm{c}$ are the (cyclic, symmetric, degree-odd) closed-string products on a sphere, starting with the closed-string BRST operator $l_1 = Q_\mathrm{c}$. Finally, $g_\mathrm{s}$ denotes the string coupling constant and $Z_\mathrm{sphere}$ the sphere partition function, which is characteristic to the chosen closed-string background. Requiring that the action \eqref{eq:CSFTAction} solves the classical BV master equation 
\begin{align}
    \frac{1}{2}\big(S_\text{sphere},S_\text{sphere}\big)=0
\end{align}
is equivalent to demanding that the products $l_k$ satisfy the $L_\infty$ relations
\begin{align}
    \sum_{k=1}^r l_k l_{r+1-k}=0\,.\label{eq:Linfty}
\end{align}
Following the Feynman rules dictated by the action \eqref{eq:CSFTAction}, one can compute tree-level on-shell closed-string amplitudes around the chosen background. On the other hand, varying this action with respect to $\Phi$ yields the equation of motion
\begin{align}
    \sum_{k=1}^\infty \frac{1}{k!} l_k(\Phi^{\wedge k})=0\,.\label{eq:CSFTEOM}
\end{align}
A solution $\Phi^\ast$ of \eqref{eq:CSFTEOM} should then be thought of as representing a new consistent classical closed-string background described in terms of a new worldsheet theory $\mathrm{CFT}^\ast$ and, in general, a new value $g_\mathrm{s}^\ast$ of the string coupling constant. By the token of the background-independence conjecture \cite{Sen:1993mh,Sen:1993kb}, 
on-shell closed-string amplitudes on this new background can be computed by expanding the action \eqref{eq:CSFTAction} in small fluctuations around $\Phi^\ast$ and applying a suitable field redefinition. 

\subsubsection{Sphere-disk open-closed vertices}

Enlarging our scope to include worldsheets with boundaries and arbitrary genus, one can formulate a consistent field theory of a degree-even open-string field $\Psi\in\mathcal{H}_\mathrm{o}$ (which is a linear combination of states in the boundary spectrum of a combined matter and ghost BCFT given by a boundary state $\|B\rangle\!\rangle$) and a degree-even closed-string field $\Phi\in\mathcal{H}_\mathrm{c}$, whose action $S_{\mathrm{oc}}[\Phi,\Psi]$ satisfies the quantum BV master equation \cite{Zwiebach:1990qj,Zwiebach:1997fe}
\begin{align}
 \frac{1}{2}\big(S_\text{oc},S_\text{oc}\big)+\Delta S_\text{oc}=0\,.\label{eq:OCBV}
\end{align}
The action $S_\mathrm{oc}$ is therefore well-suited for a complete (quantum) treatment of a backreacting system of D-branes, encoded in the boundary state $\| B\rangle\!\rangle$, on a given closed-string background specified by a worldsheet CFT and string coupling constant $g_\mathrm{s}$ \cite{Maccaferri:2023gcg,Maccaferri:2023gof}.
Expressing $S_\text{oc}$ as a sum over worldsheet topologies (parametrized by the genus $g$ and boundary number $b$), it contains the sphere part \eqref{eq:CSFTAction}, as well as the disk part
\begin{align}
    S_\mathrm{disk}[\Phi,\Psi] &= Z_\mathrm{disk}-\frac{1}{g_\mathrm{s}}\sum_{k=0}^\infty \frac{1}{(k+1)!} \omega_\mathrm{c}(\Phi,l_{k,0}(\Phi^{\wedge k}))+\nonumber\\
    &\hspace{4cm}-\frac{1}{g_\mathrm{s}}\sum_{k=0}^\infty \sum_{l=0}^\infty \frac{1}{k!}\frac{1}{l+1}\omega_\mathrm{o}\big(\Psi,m_{k,l}(\Phi^{\wedge k};\Psi^{\otimes l})\big)\,,\label{eq:DiskAction}
\end{align}
which are then followed by other (higher-$g$ and higher-$b$) contributions. The new ingredients introduced in \eqref{eq:DiskAction} include the open-string symplectic form $\omega_\mathrm{o}:\mathcal{H}_\mathrm{o}\otimes \mathcal{H}_\mathrm{o}\longrightarrow \mathbb{C}$ (defined in terms of the BPZ product on $\mathcal{H}_\mathrm{o}$), the (cyclic, degree-odd, symmetric) closed-string valued disk products $l_{k,0}:(\mathcal{H}_\mathrm{c})^{\otimes k}\longrightarrow \mathcal{H}_\mathrm{c}$, the open-string valued degree-odd products $m_{k,l}: (\mathcal{H}_\mathrm{c})^{\otimes k}\otimes (\mathcal{H}_\mathrm{o})^{\otimes l} \longrightarrow \mathcal{H}_\mathrm{o}$ and the disk partition function 
\begin{align}
Z_\mathrm{disk} = -\frac{1}{2\pi^2} \frac{g}{g_\mathrm{s}}\,,
\end{align}
where $g=\langle 0\| B\rangle\!\rangle$ denotes the $g$-function of the consistent conformal boundary state $\|B\rangle\!\rangle$.
Furthermore, as a consequence of the (classical) criticality of the vacuum $\Phi=\Psi=0$, we have $m_{0,0}=0$. The remaining products $m_{k,l}$ are symmetric in their closed-string slots and cyclic in their open-string slots. For $k>0$, the corresponding disk string vertices can be equivalently parametrized in terms of the closed-string valued disk products $l_{k,l}:(\mathcal{H}_\mathrm{c})^{\otimes k}\otimes (\mathcal{H}_\mathrm{o})^{\otimes l} \longrightarrow \mathcal{H}_\mathrm{c}$. The transition between the two descriptions is facilitated by the relation
\begin{align}
    \omega_\mathrm{o}\big(\Psi,m_{k,l}(\Phi^{\wedge k};\Psi^{\otimes l})\big) = \omega_\mathrm{c}\big(\Phi,l_{k-1,l+1}(\Phi^{\wedge (k-1)};\Psi^{\otimes (l+1)})\big)\,.
\end{align}
Altogether, the BV master equation \eqref{eq:OCBV} implies the \emph{Sphere-Disk Homotopy Algebra (SDHA)} which, on top of the $L_\infty$ relations \eqref{eq:Linfty}, includes the homotopy relations
\begin{subequations}
    \begin{align}
        \sum_{k=1}^r [l_k,l_{r-k,0}] +\sum_{k=1}^{r-1} l_{k-1,1}m_{r-k,0}&=0\,,\\
        \sum_{k=1}^r m_{k,s-1} l_{r+1-k}+ \sum_{k=0}^r\sum_{n=1}^s m_{k,n}m_{r-k,s-n}&=0\,.\label{eq:OCHA}
    \end{align}
\end{subequations}
Notice that in \eqref{eq:OCHA}, one can recognize the Kajiura-Stasheff Open-Closed Homotopy Algebra (OCHA) \cite{Kajiura:2004xu, Kajiura:2005sn}.  See \cite{Munster:2011ij,Maccaferri:2023gcg,Maccaferri:2023gof} for an analysis of the full algebraic structure of quantum open-closed SFT action vertices.

\subsubsection{Open-string field theory on a classical closed-string background}

Restricting the closed string field $\Phi$ to satisfy the classical closed-string equation of motion \eqref{eq:CSFTEOM}, one can show that the action
\begin{align}
    S_{\Phi^\ast}[\Psi] = S_\mathrm{disk}[\Phi^\ast,\Psi]\label{eq:SPsi}
\end{align}
satisfies the classical open-string BV master equation
\begin{align}
\big(S_{\Phi^\ast},S_{\Phi^\ast}\big)_\mathrm{open}=0\,.\label{eq:BVopen}
\end{align}
This means that \eqref{eq:SPsi} represents a consistent \emph{classical} action for the open-string field $\Psi$ on a closed-string background configuration given by $\Psi^\ast$. In particular, one should emphasize that the action \eqref{eq:SPsi} is therefore well-suited for the description of classical dynamics of D-branes on a changing closed-string background in the \emph{probe approximation}: while $S_{\Phi^\ast}$ will be able to encompass the response of a given D-brane system to the change in the closed-string background given by the classical closed-string field configuration $\Phi^\ast$, it will not provide any information about the backreaction of the D-branes on the bulk. The vertices of the action $S_{\Phi^\ast}[\Psi]$ can be conveniently parametrized in terms of open-string valued products $\tilde{m}_{l}:(\mathcal{H}_\mathrm{o})^{\otimes l} \longrightarrow \mathcal{H}_\mathrm{o}$ defined as
\begin{align}
  \tilde{m}_{l}(\Psi^{\otimes l})  = \sum_{k=0}^\infty \frac{1}{k!}m_{k,l}((\Phi^\ast)^{\wedge k};\Psi^{\otimes l})\,.
\end{align}
One can then write the action \eqref{eq:SPsi} in the form
\begin{align}
    S_{\Phi^\ast}[\Psi] = Z_\mathrm{disk}+\Lambda_\mathrm{c}(\Phi^\ast)- \frac{1}{g_\mathrm{s}}\sum_{l=0}^\infty \omega_\mathrm{o}\big(\Psi,\tilde{m}_l(\Psi^{\otimes l})\big)\,,\label{eq:SPsim}
\end{align}
where we have isolated the contribution
\begin{align}
 \Lambda_\mathrm{c}(\Phi^\ast)=   -\frac{1}{g_\mathrm{s}}\sum_{k=0}^\infty \frac{1}{(k+1)!} \omega_\mathrm{c}(\Phi^\ast,l_{k,0}((\Phi^\ast)^{\wedge k}))
\end{align}
of purely-closed disk vertices into the \emph{constant} part of $S_{\Phi^\ast}$.
The BV relation \eqref{eq:BVopen} is then equivalent to stating that the products $\tilde{m}_l$ satisfy an $A_\infty$ algebra. Crucially, this algebra is \emph{weak}, because the product $\tilde{m}_0$ is generally non-zero and gives rise to a tadpole in \eqref{eq:SPsim}. In particular, this means that $\Psi=0$ no longer represents a critical open-string background. So as to remove the tadpole and hence restore the full open-closed criticality, one has to re-expand the action \eqref{eq:SPsim} in small fluctuations around an open-string field configuration $\Psi^\ast$, which solves the equation of motion
\begin{align}
\sum_{l=0}^\infty  \tilde{m}_{l}(\Psi^{\otimes l})  = 0\,.\label{eq:OSEOM}
\end{align}
The pair $(\Phi^\ast,\Psi^\ast)$ of string fields should then be thought of as interpolating between the original perturbative vacuum $\Phi=\Psi=0$ and a new consistent classical open-closed background described in terms of a new bulk worldsheet theory $\mathrm{CFT}^\ast$, new string coupling $g_\mathrm{s}^\ast$ and the corresponding response $\|B^\ast\rangle\!\rangle$ of the D-brane system to the change of the bulk. In order to gain some mileage on \emph{what} exactly the new background is, one has to formulate (and then, ideally, calculate) some observable quantities. To this end, we will be interested in the {constant part} of the action $S_{\Phi^\ast}[\Psi]$ expanded around $\Psi^\ast$, which is given simply by the \emph{on-shell disk action} 
\begin{align}
S_{\Phi^\ast}[\Psi^\ast]=S_\mathrm{disk}[\Phi^\ast,\Psi^\ast]=Z_\mathrm{disk}+\Lambda(\Phi^\ast,\Psi^\ast)\,,\label{eq:OnShellDisk}
\end{align}
where we have denoted
\begin{subequations}
\begin{align}
\Lambda(\Phi^\ast,\Psi^\ast) &=  \Lambda_\mathrm{c}(\Phi^\ast)- \frac{1}{g_\mathrm{s}}\sum_{l=0}^\infty \omega_\mathrm{o}\big(\Psi^\ast,\tilde{m}_l((\Psi^\ast)^{\otimes l})\big)\\
&=-\frac{1}{g_\mathrm{s}}\sum_{k=0}^\infty \frac{1}{(k+1)!} \omega_\mathrm{c}(\Phi^\ast,l_{k,0}((\Phi^\ast)^{\wedge k}))+\nonumber\\
&\hspace{4cm}-\frac{1}{g_\mathrm{s}}\sum_{k=0}^\infty \sum_{l=0}^\infty \frac{1}{k!}\frac{1}{l+1}\omega_\mathrm{o}\big(\Psi^\ast,m_{k,l}((\Phi^\ast)^{\wedge k};(\Psi^\ast)^{\otimes l})\big)\,.\label{eq:Lam}
\end{align}
\end{subequations}
Strong evidence for the observability of the on-shell disk action \eqref{eq:OnShellDisk} is provided by the fact that the quantity $\Lambda(\Phi^\ast,\Psi^\ast)$ can be shown \cite{cosmo} to be invariant not only with respect to the (weak) $A_\infty$ gauge-variation of the open-string solution $\Psi^\ast$, but also with respect to the $L_\infty$ gauge-variation of the closed string solution $\Phi^\ast$.
Following the spirit of background independence, one is then led to conjecture that the disk action $S_\mathrm{disk}[\Phi^\ast,\Psi^\ast]$
should be identified as the disk partition function $Z_\mathrm{disk}^\ast$ of the new open-closed background $(\mathrm{CFT}^\ast,g_\mathrm{s}^\ast,\|B^\ast\rangle\!\rangle)$. Expressing the original, as well as the new disk partition function in terms of the $g$-function and the string coupling constant, we finally arrive at the \emph{generalised Sen's conjecture}
\begin{align}
     -\frac{1}{2\pi^2} \bigg(\frac{g^\ast}{g_\mathrm{s}^\ast}-\frac{g}{g_\mathrm{s}}\bigg) = \Lambda(\Phi^\ast,\Psi^\ast)\,.\label{eq:GenSenConj}
\end{align}
In the special class of cases when the closed-string background does \emph{not} undergo any change (that is $\Phi^\ast=0$), one has to put $g_\mathrm{s}^\ast =g_\mathrm{s}$ while, so that the l.h.s.\ of \eqref{eq:GenSenConj} is directly proportional to the change in the $g$-function of the D-brane system. Since, at the same time, the r.h.s.\ of \eqref{eq:GenSenConj} then reduces to the on-shell OSFT action, one recovers the classical Sen's conjecture. In the more general case when $\Phi^\ast\neq 0$, the conjecture \eqref{eq:GenSenConj} not only provides a way of tracking the change in the $g$-function of probe D-branes as the bulk is deformed, but it also represents an indirect method of measuring any potential changes in the string coupling constant: as the new value $g_\mathrm{s}^\ast$ should depend neither on the choice of the initial boundary state $\|B\rangle\!\rangle$, nor on the open-string solution $\Psi^\ast$, one should be able to isolate the change in the string coupling constant by analyzing the r.h.s.\ of \eqref{eq:GenSenConj} for a number of distinct open-string backgrounds and assuming that one \emph{knows} the expected change in the $g$-function for some of these. This demonstrates how coupling the bulk closed-string theory to probe D-branes may be beneficial for providing observables which help with keeping track of pure closed-string dynamics.

\subsection{Bulk marginal deformations}

The formal statement of the generalised Sen's conjecture can be illustrated by taking the closed-string solution $\Phi^\ast$ to represent a marginal deformation.

\subsubsection{Perturbative solutions \texorpdfstring{$\Phi^\ast(\mu)$}{TEXT} and \texorpdfstring{$\Psi^\ast(\mu)$}{TEXT}}

We will consider an exactly marginal deformation of the closed-string background which, at the leading order of the (continuous) deformation parameter $\mu$, is given by an on-shell ghost-number 2 operator $V$. We will mostly focus on the cases when $V = c\bar{c}\mathbb{V}$, where $\mathbb{V}$ is a matter primary with dimensions $(h,\bar{h})=(1,1)$. The exact marginality of the deformation needs to be guaranteed order by order in $\mu$ \cite{cosmo}. For instance, at second order, this entails requiring that
\begin{align}
    P_0^+ l_2(\mu V,\mu V) =0 \label{eq:BulkProj}
\end{align}
up to $Q_\mathrm{c}$-exact terms. The full unobstructed solution $\Phi^\ast(\mu)$ can then be expanded as
\begin{align}
      \Phi^\ast(\mu) &= \mu V - \frac{1}{2!}\frac{b_0^+}{L_0^+}\bar{P}_0^+ l_2(\mu V, \mu V) +\mathcal{O}(\mu^3)\,.\label{eq:ClSol}
\end{align}
Furthermore, let us assume that given the exactly marginal deformation $\Phi^\ast(\mu)$, one can find a corresponding solution $\Psi^\ast(\mu)$ to the open-string equation of motion \eqref{eq:OSEOM} which remains perturbatively close to the original background at $\mu=0$. This again necessitates clearing possible cohomological obstructions order by order in $\mu$, starting with
\begin{align}
    P_0 m_{1,0}(\mu V)=0\label{eq:BoundaryProj}
\end{align}
up to $Q_\mathrm{o}$-exact terms. Failure of these would signify an inability of the D-brane system at hand to adapt to the marginal deformation which we are turning on in the bulk. Instead, one would then have to search for a non-perturbative solution to \eqref{eq:OSEOM} in order to restore criticality of the background. In the absence of any obstructions, the open-string solution $\Psi^\ast(\mu)$ would be expanded in $\mu$ as
\begin{align}
    \Psi^\ast(\mu)&= -\frac{b_0}{L_0}\bar{P}_0 m_{1,0}(\mu V)+\mathcal{O}(\mu^2)\,.\label{eq:OpSol}
\end{align}

\subsubsection{The deformed disk partition function}

Substituting the solutions \eqref{eq:ClSol} and \eqref{eq:OpSol} into \eqref{eq:Lam} yields a $\mu$-expansion of the change in the disk partition function. Combining this with the statement \eqref{eq:GenSenConj} of the generalised Sen's conjecture, one obtains 
\begin{align}
     &\frac{1}{2\pi^2} \bigg[\frac{g^\ast(\mu)}{g}\bigg(\frac{g_\mathrm{s}^\ast(\mu)}{g_\mathrm{s}}\bigg)^{-1}-1\bigg] =\nonumber\\
     &\hspace{1cm}=\frac{1}{g}\omega_\mathrm{c}\big(\mu V,l_{0,0}\big)+\frac{1}{2}\bigg[\frac{1}{g}\omega_\mathrm{c}\bigg(l_{0,0},\frac{b_0^+}{L_0^+}\bar{P}_0^+ l_2(\mu V ,\mu V)\bigg)+\nonumber\\
     &\hspace{3cm}+\frac{1}{g}\omega_\mathrm{c}\big(\mu V, l_{1,0}(\mu V)\big)+\frac{1}{g}\omega_\mathrm{o}\bigg(m_{1,0}(\mu V),\frac{b_0}{L_0}\bar{P}_0 m_{1,0}(\mu V)\bigg)\bigg]+\mathcal{O}(\mu^3)\,.\label{eq:CosmoExp}
\end{align}
To evaluate the r.h.s.\ of \eqref{eq:CosmoExp}, it is practical to make concrete assumptions about the form of the SFT vertices. See \cite{cosmo} for a particular choice of vertices which are based on $SL(2,\mathbb{C})$ maps and which will be also put to use in later sections of this paper when discussing relevant deformations. One should nonetheless keep in mind that as the disk action is an observable quantity, the endpoint of calculating the r.h.s.\ of \eqref{eq:CosmoExp} should be manifestly independent of any off-shell SFT data (such as the local coordinate maps around punctures and their parameters). This serves as a useful consistency check. Following the steps outlined in \cite{cosmo}, one can then arrive at the expression
\begin{align}
   &\frac{g^\ast(\mu)}{g}\bigg(\frac{g_\mathrm{s}^\ast(\mu)}{g_\mathrm{s}}\bigg)^{-1} = 1+ 2\,\frac{1}{g}\big\langle \mu\mathbb{V}(i,\bar{i})\big\rangle_{\mathrm{UHP}}+ \int\limits_{\substack{0\\ \mathrm{reg.}}}^1 ds\,(s^2-1)\,\frac{1}{g}\big\langle \mu\mathbb{V}(i,\bar{i})\,\mu\mathbb{V}(is,\bar{is})\big\rangle_\mathrm{UHP}+\mathcal{O}(\mu^3)\label{eq:DiskActionMarg}
\end{align}
for the evolution of the worldhseet disk partition function with $\mu$. The integral in the second term runs from $s=0$ (open-string degeneration) to $s=1$ (closed-string degeneration) and needs to be regulated at both endpoints so as to avoid divergences coming from the propagation of the open- and closed-string tachyons. In particular, we define \cite{Maccaferri:2021ksp,cosmo} 
\begin{align}
\int\limits_{\substack{0\\ \mathrm{reg.}}}^1  ds\, = \lim_{\substack{\epsilon_\mathrm{c}\to 0\\ \epsilon_\mathrm{o}\to 0}}\,\bigg[ \int_0^a ds\, s^{\epsilon_\mathrm{o}}+\int_a^1 ds\, \bigg(\frac{1-\sqrt{s}}{1+\sqrt{s}}\bigg)^{\epsilon_\mathrm{c}}\bigg]\,,\label{eq:RegInt}
\end{align}
where the parameter $a\in (0,1)$ serves to separate the two degeneration regions (and can be related to the open-string stub parameter $\lambda_\mathrm{o}$ as $a = 1/\lambda_\mathrm{o}^2$). The result of the integration is however always independent of $a$, as can be readily checked by differentiating with respect to it. Also note that the function appearing in the second term of \eqref{eq:RegInt} regulating the closed-string collision can be naturally interpreted in terms of the radial coordinate $y$ on the disk, so that near the closed-string degeneration ($y=0$), the tachyon-divergence is regulated with $y^{\epsilon_\mathrm{c}}$, as appropriate. Finally, note that the integral is not endangered by any divergences due to propagation of massless modes as these are absent precisely by virtue of the conditions \eqref{eq:BulkProj} (closed-string channel) and \eqref{eq:BoundaryProj} (open-string channel).

\subsection{Example: navigating the Narain moduli space}

As an illustration, we will now consider computing the on-shell disk action in the case when the on-shell closed-string field $\Phi^\ast$ represents a finite deformation of the Narain modulus $E_{\mu\nu} = G_{\mu\nu}+B_{\mu\nu}$ of a toroidal compactification of $d$ free bosons. We will assume that our probe D-brane is represented by an elementary matter boundary state which satisfies the Ishibashi conditions
\begin{align}
    \big[(\alpha_{n}^\mathrm{L})_{\mu}+\tensor{\Omega}{_\mu^\nu}(\alpha_{-n}^\mathrm{R})_{\nu}\big]\|B\rangle\!\rangle =0\,,\label{eq:Ishibashi}
\end{align}
where the gluing automorphism matrix $\Omega$ is constrained to satisfy $\Omega G \Omega^{T}=G$. Such boundary state is built up from $U(1)^d$ Ishibashi states, which are labelled by the left- and right-moving momenta
\begin{subequations}
\begin{align}
    k^\mathrm{L} &= k +E w\,,\\
    k^\mathrm{R} &= k -E^T w\,,
\end{align}
\end{subequations}
where the momentum $k_\mu$ and winding $w^\nu$ are both $\mathbb{Z}^d$-valued. Specializing the relation \eqref{eq:Ishibashi} for the zero-modes $n=0$, one learns that $k$ and $w$ need to be restricted so that
\begin{align}
    (1+\Omega)k+(E-\Omega E^T)w=0\,.\label{eq:kwOmega}
\end{align}
Requiring that \eqref{eq:kwOmega} is solved by a $d$-dimensional sublattice of $\mathbb{Z}^d\oplus \mathbb{Z}^d$ imposes further restrictions on $\Omega$, which in fact make the set of allowed gluing matrices $\Omega$ discrete. 

\subsubsection{BCFT expectation}

Let us first discuss what change in the $g$-function of a Narain boundary state one should expect (based on pure BCFT considerations) as we deform the Narain modulus by sending $E\to E+\epsilon_\sigma$. In conformal perturbation theory, this change can be implemented by the marginal deformation
\begin{align}
    S[X;E]\longrightarrow S[X;E]+ \frac{1}{4\pi} \int d^2 z\, \mathbb{V}^{\epsilon_\sigma}(z,\bar{z})\,,
\end{align}
where $S[X;E]$ is the action of $d$ free bosons $X^{\mu}$ compactified on lattice with Narain modulus $E_{\mu\nu}$ and 
\begin{align}
   \mathbb{V}^{\epsilon_\sigma}(z,\bar{z})  =(\epsilon_\sigma)_{\mu\nu} \,\p X^\mu(z)\, \bar{\p} X^\nu(\bar{z})\label{eq:DefOpNarain}
\end{align}
is the deforming operator.
To first order in $\epsilon_\sigma$, a one-loop calculation in conformal perturbation theory reveals that the $g$ function changes as
\begin{align}
    \frac{\delta^{(1)}g}{g} = \frac{1}{4}\Tr\big[\Omega^T G^{-1}\epsilon_\sigma\big]\,.\label{eq:res1}
\end{align}
Furthermore, varying the relation \eqref{eq:kwOmega} with respect to $E$, it is straightforward to realise that
\begin{align}
    \delta^{(1)} \Omega  = \frac{1}{2}(\epsilon_\sigma - \Omega\, \epsilon_\sigma^T)G^{-1}(1+\Omega)\,.\label{eq:res2}
\end{align}
Given the results \eqref{eq:res1} and \eqref{eq:res2}, one can straightforwardly find the second variation $\delta^{(2)}g$ of $g$ with respect to $E$. Combining this with the first variation \eqref{eq:res1}, we learn that the finite change $\Delta g$ of the $g$-function can be expanded to second order in $\epsilon_\sigma$ as
\begin{subequations}
\begin{align}
   1+ \frac{\Delta g}{g} &= 1+\frac{\delta^{(1)}g}{g}+\frac{1}{2}\frac{\delta^{(2)}g}{g}+\mathcal{O}(\epsilon_\sigma^3) \\
    &= 1+ \frac{1}{4}\Tr\big[\Omega^T G^{-1}\epsilon_\sigma\big]+\frac{1}{32}\Tr\big[\Omega^T G^{-1}\epsilon_\sigma\big]^2 - \frac{1}{16}\Tr\big[(\Omega^T G^{-1}\epsilon_\sigma)^2\big]+\nonumber\\
    &\hspace{2cm}-\frac{1}{8}\Tr\big[\Omega^T G^{-1}\epsilon_\sigma  G^{-1}\epsilon_\sigma\big]+\frac{1}{16}\Tr\big[ G^{-1}\epsilon_\sigma^T  G^{-1}\epsilon_\sigma\big]+\mathcal{O}(\epsilon_\sigma^3)\,.\label{eq:DelgCFT}
\end{align}
\end{subequations}
Notice that all terms in \eqref{eq:DelgCFT} depend on the gluing matrix $\Omega$ (and thus on the boundary condition) except for the last one, which is proportional to the two-point function coefficient $C^{\epsilon_\sigma}_{\mathbb{V}\mathbb{V}\mathbbm{1}}$ of the deforming operator $\mathbb{V}^{\epsilon_\sigma}(z,\bar{z})$ on the sphere. Indeed, substituting from \eqref{eq:DefOpNarain}, one has
\begin{align}
    \big\langle \mathbb{V}^{\epsilon_\sigma}(z,\bar{z})\,\mathbb{V}^{\epsilon_\sigma}(w,\bar{w})\big\rangle_{\mathbb{C}} = \frac{C_{\mathbb{V}\mathbb{V}\mathbbm{1}}^{\epsilon_\sigma}}{|z-w|^4}\,,
\end{align}
where
\begin{align}
C_{\mathbb{V}\mathbb{V}\mathbbm{1}}^{\epsilon_\sigma} = \frac{1}{4}\Tr\big[ G^{-1}\epsilon_\sigma^T  G^{-1}\epsilon_\sigma\big]\,.\label{eq:CVV1}
\end{align}

\subsubsection{SFT calculation}

We will now implement the marginal deformation which sends $E$ to $E+\epsilon_\sigma$ as a classical solution (of the form \eqref{eq:ClSol}) in closed string field theory. To this end, we put
\begin{align}
    \mu \mathbb{V}(z,\bar{z})\equiv \mathbb{V}^\epsilon(z,\bar{z}) \equiv \epsilon_{\mu\nu}\, \p X^\mu (z)\,\bar{\p} X^\nu(\bar{z})\,.
\end{align}
It is not difficult to check that for such a marginal state, the condition \eqref{eq:BulkProj} holds, which in turn guarantees exact marginality of the deformation to second order in $\epsilon$.
First, one can compute the change in conformal weights of the bulk momentum plane waves induced by the solution \eqref{eq:ClSol}. This calculation proceeds by analyzing the cohomology of the BRST operator shifted around the classical solution. Comparing the result with the shift in conformal weights which is expected upon deforming the Narain modulus $E$ by sending $E\to E+\epsilon_\sigma$, we can fix the relation between the SFT deformation parameter $\epsilon$ and the CFT deformation $\epsilon^\sigma$ to be \cite{cosmo}
\begin{align}
    \epsilon = \epsilon_\sigma - \frac{1}{2} \epsilon_\sigma G^{-1}\epsilon_\sigma + \mathcal{O}(\epsilon_\sigma^3)\,.\label{eq:Redefinition}
\end{align}
Second, considering any D$p$-brane described by a gluing matrix $\Omega$, we would like to evaluate the disk action \eqref{eq:DiskActionMarg} of the open-string solution \eqref{eq:OpSol} computed on a classical closed-string background given by the solution \eqref{eq:ClSol}. This is a healthy solution, as one can readily verify that the condition \eqref{eq:BoundaryProj} holds for any $\Omega$. It describes how the D-brane adapts as we perturb the Narain modulus of the bulk CFT.
Evaluating the relevant UHP correlators, we obtain
\begin{subequations}
\label{eq:Correlators}
    \begin{align}
       \frac{1}{g} \big\langle \mathbb{V}^\epsilon(i,\bar{i})\big\rangle_{\mathrm{UHP}} &=\frac{1}{8}\Tr[\epsilon\Omega^T G^{-1}]\,,\\
     \frac{1}{g}\big\langle \mathbb{V}^\epsilon(i,\bar{i})\,\mathbb{V}^\epsilon(is,\bar{is})\big\rangle_\mathrm{UHP}   &=\frac{1}{4}\bigg(\frac{\Tr[\epsilon\Omega^T G^{-1}]^2}{16s^2}+\frac{\Tr [\epsilon G^{-1}\epsilon^T G^{-1}]}{(1-s)^4}+\frac{\Tr[\epsilon\Omega^T G^{-1}\epsilon\Omega^T G^{-1}]}{(1+s)^4}\bigg)\,.
    \end{align}
\end{subequations}
Using the prescription \eqref{eq:RegInt}, we can also evaluate
\begin{subequations}
\label{eq:Integrals}
    \begin{align}
        \int\limits_{\substack{0\\ \mathrm{reg.}}}^1  ds\,\frac{s^2-1}{16s^2}&=+\frac{1}{8}\,,\\
        \int\limits_{\substack{0\\ \mathrm{reg.}}}^1  ds\,\frac{s+1}{(s-1)^3}&=+\frac{1}{8}\,,\\
        \int\limits_{\substack{0\\ \mathrm{reg.}}}^1  ds\,\frac{s-1}{(s+1)^3}&=-\frac{1}{4}\,.
    \end{align}
\end{subequations}
Finally, substituting the results \eqref{eq:Redefinition}, \eqref{eq:Correlators} and \eqref{eq:Integrals} into the expression \eqref{eq:DiskActionMarg} for the disk action, we obtain
\begin{align}
       &\bigg(1+ \frac{\Delta g}{g}\bigg)\bigg(1+ \frac{\Delta g_\mathrm{s}}{g_\mathrm{s}}\bigg)^{-1} =\nonumber\\
       &\hspace{0.1cm}=\bigg(1+\frac{1}{4}\Tr[\epsilon_\sigma\Omega^T G^{-1}]+\frac{1}{32}\Tr[\epsilon_\sigma\Omega^T G^{-1}]^2-\frac{1}{16}\Tr[(\epsilon_\sigma\Omega^T G^{-1})^{2}]+\nonumber\\
       &\hspace{0.8cm}-\frac{1}{8}\Tr[  \epsilon_\sigma G^{-1}\epsilon_\sigma\Omega^T G^{-1}]+\frac{1}{16}\Tr [\epsilon_\sigma G^{-1}\epsilon_\sigma^T G^{-1}]\bigg)\bigg(1+\frac{1}{32}\Tr [\epsilon_\sigma G^{-1}\epsilon_\sigma^T G^{-1}]\bigg)^{-1}+\mathcal{O}(\epsilon_\sigma^3)\,.\label{eq:DiskActionNarain}
\end{align}
Quite non-trivially, on the r.h.s. of \eqref{eq:DiskActionNarain} we recover the expected response \eqref{eq:DelgCFT} of the $g$-function of the D$p$-brane boundary state to the bulk deformation $E\to E+\varepsilon_\sigma$. This is multiplied by a universal factor which is independent of $\Omega$. Attributing this factor to a change $\Delta g_\mathrm{s}$ in the string coupling constant $g_\mathrm{s}$ which is induced by the closed-string solution \eqref{eq:ClSol}, we can write
\begin{align}
   1+ \frac{\Delta g_\mathrm{s}}{g_\mathrm{s}}= 1+\frac{1}{8}C_{\mathbb{V}\mathbb{V}\mathbbm{1}}^{\epsilon_\sigma}+\mathcal{O}(\epsilon_\sigma^3)\,,\label{narain-gs}
\end{align}
where the two-point function coefficient $C_{\mathbb{V}\mathbb{V}\mathbbm{1}}^{\epsilon_\sigma}$ was computed in \eqref{eq:CVV1}.


\section{Short bulk RG flows with boundaries}\label{sec:3}
In this section, we will apply the open-closed SFT \ch{framework} to investigate how relevant bulk CFT perturbations affect boundary states. \ch{In particular, we will construct the classical open SFT solutions that describe D-branes after the closed-string background has undergone a `short' RG flow.} \ch{The possibility of studying such closed-string perturbations} via SFT was considered long ago by Mukherji and Sen \cite{Mukherji:1991tb} and more recently in \cite{Scheinpflug:2023lfn} and \cite{Mazel:2024alu}. Here we would like to extend this analysis to the presence of world-sheet boundaries representing D-branes.

We will begin our exposition by describing the pure 2d CFT setup and briefly reviewing the results of \cite{Scheinpflug:2023lfn}. We will then proceed with an explicit construction of the open SFT solutions.


\subsection{Conformal perturbation theory description}\label{sec: 3.1}
Let us start with an overview of conformal perturbation theory (CPT) for a theory ${\rm CFT}_{0}$ with boundary \ch{described by a boundary state $\| B_0\rangle\!\rangle$}. In general, bulk deformations of 2d CFTs with boundaries can induce boundary RG flows. To write down the RG equations, we consider \ch{a generic bulk and boundary deformation} given in terms of \ch{bulk primary fields $\mathbb{O}_k$ with scaling dimensions $\Delta_{k}$ and boundary primary fields $\mathbbm{o}_{l}$ with scaling dimensions $h_{l}$ as} 
    \begin{align}
          S_{{\rm CFT}_{0}}\to  S_{{\rm CFT}_{0}}+\sum_{i} \lambda_{i}\epsilon^{\Delta_i-2}\int d^2 z \, \mathbb{O}_i (z,\bar{z})+\sum_{j} \mu_{j}\epsilon^{h_j-1}\int dx \,\mathbbm{o}_i (x)\,,
    \end{align}
where $\lambda_{i}$ and $\mu_{j}$ are dimensionless coupling constants, and $\epsilon$ is the length scale used as a natural cutoff to treat the contact \ch{divergences} arising in correlators of the perturbed theory. Following the usual CPT approach (see, for example,  \cite{Fredenhagen:2006dn}), we obtain the following RG equations
\begin{subequations}
\begin{align}
\label{eq:RGeq1}
    &\frac{d\lambda_k}{d\log(\epsilon)}=(2-\Delta_k)\lambda_k+\sum_{ij}\pi C_{ijk}\lambda_i\lambda_j+\mathcal{O}(\lambda^3)\,,\\
    \label{eq:RGeq2}
    &\frac{d\mu_l}{d\log(\epsilon)}=(1-h_l)\mu_l+\sum_{i}\frac{1}{2}\frac{B_{il}}{g}\lambda_i+\mathcal{O}(\lambda \mu)+\mathcal{O}(\lambda^2)+\mathcal{O}(\mu^2) \,,
\end{align}
\end{subequations}
where $g$ is the $g$-function of the initial BCFT and $C_{ijk}$ and $B_{il}/g$ are the OPE coefficients
\begin{subequations}
\begin{align}
    &\mathbb{O}_{i}(z,\bar{z})\,\mathbb{O}_{j}(w,\bar{w})=\sum_{k}\frac{C_{ijk}}{\vert z-w\vert^{\Delta_i+\Delta_j-\Delta_k}}\mathbb{O}_{k}(w,\bar{w})+\dots\,,\\
    &\mathbb{O}_{i}(x+is,x-is)=\sum_{l}\frac{1}{(2s)^{\Delta_i-h_l}}\frac{B_{il}}{g}\mathbbm{o}_l(x)+\dots \,.
\end{align}
\end{subequations}
In particular, we will consider a perturbation triggered by a nearly marginal primary bulk field $\mathbb{V}(z,\bar{z})$ 
\ch{which has conformal dimensions $(1-y,1-y)$ with $y$ small and positive. We will also assume that the corresponding \ch{Virasoro representation} satisfies the fusion rule}
\begin{equation}
    \mathbb{V}\times \mathbb{V}=1+\mathbb{V}+\dots\,,\label{eq:fusionShort}
\end{equation}
where the dots represent possible irrelevant contributions. The bulk and the bulk-boundary OPEs of $\mathbb{V}$ can then be written as
\begin{align}
\mathbb{V}(z,\bar{z})\mathbb{V}(w,\bar{w})&\stackrel{z\to w}{\sim}\frac{1}{\vert z-w\vert^{4(1-y)}}+\frac{1}{\vert z-w\vert^{2(1-y)}}\tensor{C}{_{\mathbb{VVV}}}\mathbb{V}(w,\bar{w})+{\rm reg}\,,\\
    \mathbb{V}(x+is,x-is)&\stackrel{s\to 0}{\sim}\frac{1}{(2s)^{2(1-y)}} \frac{\tensor{B}{_{\mathbb{V}\mathbbm{1}}}}{g}+\frac{1}{(2s)^{1-y}}\frac{\tensor{B}{_{\mathbb{V}\mathbbm{v}}}}{g}\mathbbm{v}(x)+{\rm reg}\,,
\end{align}
where we normalised the nearly marginal field as $\langle \mathbb{V} \vert \mathbb{V} \rangle =\tensor{C}{_{\mathbb{V}\mathbb{V}\mathbbm{1}}}= 1$. \ch{We have also introduced the boundary field $\mathbbm{v}$ of dimension $h = 1 - y$ transforming in the same Virasoro representation as the deforming bulk field $\mathbb{V}$. Note that sometimes, it may happen that this representation is absent from the spectrum of the boundary fields as determined by the Cardy condition (annulus crossing equation). In such cases, the OPE coefficient $\tensor{B}{_{\mathbb{V}\mathbbm{v}}}/g $ has to vanish.} 

Under these assumptions the RG  equations \eqref{eq:RGeq1} and \eqref{eq:RGeq2} can be rewritten as 
\begin{subequations}
\begin{align}
\label{eq:RGeqttau1}
    \frac{dt}{d\log(\epsilon)}&=2yt -\tensor{C}{_{\mathbb{VVV}}}t^2+\mathcal{O}(t^3)\,,\\
    \label{eq:RGeqttau2}
    \frac{d\tau}{d\log(\epsilon)}&=y \tau +\frac{\tensor{B}{_{\mathbb{V}\mathbbm{v}}}}{g}\frac{t}{2}+\mathcal{O}(t \tau)+\mathcal{O}(t^2)+\mathcal{O}(\tau^2)\,,
\end{align}
\end{subequations}
where $\tau\coloneqq -\mu_{\mathbbm{v}} \pi$. In turn, from \eqref{eq:RGeqttau1} and \eqref{eq:RGeqttau2} we can generally read off the fixed points at leading order in $y$ as
\begin{subequations}
\begin{align}
    t^{\ast}(y)&=\frac{2y}{\tensor{C}{_{\mathbb{VVV}}}}+\mathcal{O}(y^2)\,,\label{eq:fpBulk}\\
    \tau^{\ast}(y)&=-\frac{1}{\tensor{C}{_{\mathbb{VVV}}}}\frac{\tensor{B}{_{\mathbb{V}\mathbbm{v}}}}{g}+\mathcal{O}(y)\,.
\end{align}
\end{subequations}
Looking at the first line, we can notice that $y$ adjusts the length of the bulk RG flow by making it short in the $y\to 0$ limit, in the sense that we have a perturbative bulk RG fixed point $t^{\ast}(y)$. \ch{On the other hand,} in the case with boundaries, we have a perturbative fixed point (with $\tau^{\ast}(0)=0$) only if 
\begin{equation}
\label{eq:assumption}
    \frac{\tensor{B}{_{\mathbb{V}\mathbbm{v}}}}{g}\biggr\vert_{y=0}=0\,.
\end{equation}
Therefore, to have a short bulk RG flow with boundaries, we must work under the assumption $\tensor{B}{_{\mathbb{V}\mathbbm{v}}}/g \stackrel{y \to 0}{\sim} \mathcal{O}(y)$. \ch{This includes the cases when the bulk-broundary OPE coefficient $\tensor{B}{_{\mathbb{V}\mathbbm{v}}}/g $ vanishes exactly. In such situations, one has a choice of not triggering the boundary RG flow at all. Finally, in the cases when the field $\mathbbm{v}$ is not present in the boundary spectrum, the boundary RG equation \eqref{eq:RGeqttau2} becomes vacuous and there is no boundary RG flow to be induced by the bulk deformation triggered by $\mathbb{V}$.}

\ch{As we are now going to discuss, from the SFT perspective, the bulk-boundary perturbation triggered by $\mathbb{V}$ can be described by 1.\ starting with a consistent open-closed world-sheet background incorporating the initial matter theory $\text{CFT}_0$ and boundary state $\|B_0\rangle\!\rangle$, 2.\ solving perturbatively the closed-string equation of motion for a classical solution $\Phi^\ast(y)$, 3.\ shifting the open-closed background by this solution and then, finally, 4.\ solving the tadpole-sourced open-string equation of motion to find the classical solution $\Psi^\ast(y)$ which captures the changes in the D-brane system along the RG flow.}

\subsection{Short bulk RG flows from SFT}\label{sec:3.2}

Starting with the bulk RG flows, we embed the matter \ch{theory} ${\rm CFT}_0$ with central charge $c$, into a worldsheet CFT given by the tensor product of three sectors
\begin{equation}
\label{eq:bulkbackground}
    {\rm CFT}_{\rm tot}^{(0)}={\rm CFT}_{0}\otimes {\rm CFT}_{\rm aux}\otimes {\rm CFT_{gh}}\,,
\end{equation}
where ${\rm CFT_{gh}}$ is the usual $bc$-ghost system with central charge $c_{\rm gh} = -26$ and ${\rm CFT_{aux}}$ is an auxiliary CFT with $c_{\rm aux} = 26 - c$, which ensures that the theory is critical,
\begin{equation}
    c_{\rm tot}=c+c_{\rm aux}+c_{\rm gh}=0\,.
\end{equation}
Now, our aim is to perturbatively solve the classical closed-string equation of motion \eqref{eq:CSFTEOM}. \ch{Doing so, one should find} an explicit expression for the solution $\Phi^{\ast}(y)$ which describes the critical closed-string background \ch{corresponding to fixed point \eqref{eq:fpBulk} of} the short RG flow. Since our eventual goal  will be to evaluate the \ch{the on-shell disk action \eqref{eq:OnShellDisk}} up to quadratic order, \ch{we will only be interested in finding $\Phi^{\ast}(y)$ up to second order in $y$.} In CPT, this would correspond to a rather hard two-loop computation. 
\ch{Also recall that as we are neglecting backreaction of the D-branes on the bulk, the closed-string equation of motion is unaffected by the presence of the boundaries, exactly in the spirit of the bulk RG equations \eqref{eq:RGeq1}.} 

\subsubsection{Closed-string equation of motion and obstructions}\label{sec:3.2.1}

Following the strategy outlined in \cite{Mukherji:1991tb} and \cite{Scheinpflug:2023lfn}, we introduce a projector $P$ that projects \ch{onto the space of eigenstates of the operators $L_{0}$ and $\bar{L}_{0}$ which have eigenvalues $(0,0)$ and $(-y,-y)$.} \ch{It is not hard to see that $P$} commutes with the BRST charge, \ch{that is} $[P,Q_{\rm c}]=0$. \ch{Afterwards}, we split the closed string field through the action of $P$ and $\bar{P}=1-P$ \ch{by} defining
\begin{equation}
\label{eq:splitsol}
\Phi=P\Phi+\bar{P}\Phi\coloneqq W + R\,.
\end{equation}
\ch{For the ``tachyon'' component $W$ of the string field, we then consider the ansatz}
\begin{equation}
\label{eq:ansatz}
    W = t(y) T\,,
\end{equation}
where $T=c\bar{c}\mathbb{V}$ is the nearly on-shell tachyon and $t(y)$ is a function that will be fixed by imposing the closed-string equation of motion. In particular, this function can be expanded as
\begin{align}
    &t(y)=t_1 y+ t_2 y^{2}+\mathcal{O}(y^3)\,.
\end{align}
Our aim will be to explicitly find $t_1$ and $t_2$ in terms of CFT data.

\ch{First, let us} project the \ch{full} equation of motion
\begin{equation}
    {\rm EOM}_{\rm c}[\Phi]=Q_{\rm c}\Phi+\frac{1}{2}l_{2}(\Phi,\Phi)+\frac{1}{3!}l_3(\Phi,\Phi,\Phi)+\mathcal{O}(y^4)
\end{equation}
\ch{into the two (orthogonal) subspaces given as the image of $P$ and $\bar{P}$, namely}
\begin{subequations}
\begin{align}
\label{eq:PCEOM}
    &P\,{\rm EOM}_{\rm c}[\Phi]=Q_{\rm c}W+\frac{1}{2}Pl_{2}(\Phi,\Phi)+\frac{1}{3!}Pl_3(\Phi,\Phi,\Phi)+\mathcal{O}(y^4)\,,\\
    &\bar{P}\,{\rm EOM}_{\rm c}[\Phi]=Q_{\rm c}R+\frac{1}{2}\bar{P}l_{2}(\Phi,\Phi)+\frac{1}{3!}\bar{P}l_3(\Phi,\Phi,\Phi)+\mathcal{O}(y^4)\,.
\end{align}
\end{subequations}
\ch{Since it is possible to invert $Q_\mathrm{c}$ on the image of $\bar{P}$, this splitting enables us to write down a recursive solution for $R$ in terms of $W$. At leading order, this reads (in the Siegel gauge $b_{0}^{+}=0$)}
\begin{equation}
\label{eq:R}
    R(W)=-\frac{1}{2}\frac{b_{0}^{+}}{L_{0}^{+}}\bar{P}l_2(W,W)+\mathcal{O}(y^3)\,.
\end{equation}
\ch{Plugging \eqref{eq:R} back into \eqref{eq:PCEOM}, the component of the full equation of motion which lies in the image of $P$ can be recast purely in terms of $W$ as}
\begin{equation}
\label{eq:PCEOM1}
    P\,{\rm EOM}_{\rm c}[\Phi]=Q_{\rm c}W+\frac{1}{2}Pl_{2}(W,W)-\frac{1}{2}Pl_{2}\left(W,\frac{b_{0}^{+}}{L_{0}^{+}}\bar{P}l_2(W,W)\right)+\frac{1}{3!}P l_3(W,W,W)+\mathcal{O}(y^4)\,.
\end{equation}
Looking \ch{in detail} at \ch{the r.h.s.\ of \eqref{eq:PCEOM1}}, we can notice that there is an obstruction in solving the equation of motion. Indeed, the last term \ch{turns out to} give rise to terms proportional to the anti-ghost dilaton, $c_{0}^{+}\left(c\p^2 c-\bar{c}\bar{\p}^2\bar{c}\right)$ at cubic order in $y$, \ch{which cannot be balanced by any of the remaining terms on the r.h.s.\ of \eqref{eq:PCEOM1}.} The \ch{meaning} of this obstruction is related to the fact that \ch{the deformation by $\mathbb{V}$ gives rise to an $\mathcal{O}(y^3)$ change in} the central charge \ch{of the matter CFT.} \ch{With no further provisions, this would have resulted into} a non-critical total world-sheet CFT. 

A solution to this problem was \ch{suggested already} in \cite{Mukherji:1991tb} and very recently concretised in \cite{Mazel:2024alu}. The key point is to modify the auxiliary sector of the initial \ch{worldsheet} CFT by adding a linear dilaton sector $\mathrm{CFT}_{{\rm D}_{\beta}}$. \ch{This has the} central charge $c_{\beta}=1+3\beta^2$ where $\beta$ denotes the background charge. \ch{Hence, the total worldsheet theory ${\rm CFT}_{\rm tot}^{(0)}$ now factorizes as}
\begin{equation}
\label{eq:bulkbackgroundwdilaton}
    {\rm CFT}_{\rm tot}^{(0)}={\rm CFT}_{0}\otimes {\rm CFT}_{{\rm D}_{\beta}}\otimes {\rm CFT}_{\rm aux}\otimes {\rm CFT}_{\rm gh}\,.
\end{equation}
\ch{Denoting by $Y$ the scalar field of the linear dilaton CFT, the anti-ghost dilaton can then be BRST-trivialized by the field}
 \begin{equation}
 \label{eq:theta}
     \Theta = \frac{1}{2}\left(c\p^2 c-\bar{c}\bar{\p}^2\bar{c}\right)Y+c_{0}^{+}(c\p Y- \bar{c}\bar{\p}Y)\,.
 \end{equation}
\ch{Indeed, we have}
 \begin{equation}
     Q_{\rm c }\Theta = \sqrt{\frac{\alpha^{\prime}}{2}}\beta c_{0}^{+}\left(c\p^2 c-\bar{c}\bar{\p}^2\bar{c}\right)\,.
 \end{equation}
\ch{This makes it possible} to compensate for the obstruction by introducing a term proportional to $\Theta$ in the SFT solution, namely 
\begin{equation}
    W=t(y)T+\theta(y)\Theta\,,
\end{equation}
where $\theta(y)$ is a function  which starts at the order $y^3$.
From the \ch{worldsheet} CFT viewpoint, this corresponds to implementing the RG flow both in the matter and in the linear dilaton sector. The latter will induce a change in the background charge and, therefore, in $c_{\beta}$. \ch{This, in turn, exactly compensates for the variation of the matter central charge, resulting in a total worldsheet CFT that remains critical. We therefore recover}
 \begin{equation}
     c_{\rm tot}^{\ast}=c^{\ast}+c_{\beta}^{\ast}+c_{\rm aux}+c_{\rm gh}=0\,,
 \end{equation}
where $\ast$ indicates that the corresponding quantities refer to the perturbed theory. 

\ch{Finally, having discussed the subtleties of searching for a closed-string solution which changes the matter central charge, we recall that our primary focus in this paper is on computing the disk action on the classical solution up to \emph{second} order in $y$. Hence, the third-order term $\theta(y)\Theta$ will not be of any practical interest in the remainder of our analysis.}

\subsubsection{The classical closed-string solution}\label{sec:3.2.2}
Let us now impose the equation of motion \eqref{eq:PCEOM1} and substitute the ansatz \eqref{eq:ansatz}. We obtain 
\begin{equation}
\label{eq:PCEOM2}
    Q_{\rm c}T+\frac{t(y)}{2}Pl_{2}(T,T)+\frac{t(y)^2}{3!}\left(P l_3(T,T,T)-3Pl_{2}\left(T,\frac{b_{0}^{+}}{L_{0}^{+}}\bar{P}l_2(T,T)\right)\right)+\mathcal{O}(y^3)=0\,.
\end{equation}
\ch{By calculating the symplectic form of $T$ against the l.h.s.\ of \eqref{eq:PCEOM2}, we can write}
\begin{equation}
\label{eq:PCEOM3}
    \wc\left(T,Q_{\rm c}T\right)+\frac{t(y)}{2}\wc\left(T,l_2(T,T)\right)+\frac{t(y)^2}{3!}{\cal A}_{TTTT}+\mathcal{O}(y^3)=0\,,
\end{equation}
where we defined
\begin{equation}
    {\cal A}_{TTTT}\coloneqq \wc\left(T, l_3(T,T,T)\right)+3\,\wc\left(l_2(T,T),\frac{b_{0}^{+}}{L_{0}^{+}}\bar{P}l_2(T,T)\right)\,.
    \end{equation}
The quantity $ {\cal A}_{TTTT}$ is proportional to \ch{the zero-momentum} amplitude of four slightly off-shell tachyons on the sphere.

Let us now separately describe the three terms on the l.h.s.\ of \eqref{eq:PCEOM3}. \ch{Starting with the first one, we can write}
\begin{equation}
\label{eq:TT}
    \wc(T,Q_{\rm c}T)=\langle c\Bar{c}\mathbb{V}(0,\Bar{0})\vert c_{0}^{-}(c_{0}L_{0}+\bar{c}_{0}\bar{L}_{0})\vert c\Bar{c}\mathbb{V}(0,\Bar{0})\rangle =y\,,
\end{equation}
in which we used the fact that we are working in the Siegel gauge and with operators normalized \ch{so that} $\ch{\langle 0|} c_{-1}\bar{c}_{-1} c_{0}\bar{c}_{0} c_{1}\bar{c}_{1}\ch{|0\rangle}=-1$ and $\langle \mathbb{V}\vert\mathbb{V}\rangle=1$. To evaluate the second term in \eqref{eq:PCEOM3}, we must \ch{introduce the cubic string product} $l_2$. In particular, we will use the ${\rm SL}(2,\mathbb{C})$ vertices defined in \cite{cosmo}. \ch{We will describe these} in \ch{more completeness in the following section}. For now, we will just provide the definition of $l_2$\ch{, namely} 
\begin{equation}
    \vert l_2(\Phi_1,\Phi_2)\rangle = b_{0}^{-}\delta(L_{0}^{-})f_1\circ \Phi_1(0,\Bar{0})f_2\circ \Phi_2(0,\Bar{0})\vert 0\rangle_{{\rm SL}(2,\mathbb{C})}\,,
\end{equation}
with 
\begin{equation}
    f_1(w)=\frac{1}{\lambda_{\rm c}}\frac{w-\lambda_{\rm c}}{3\lambda_{\rm c}+w}=-f_2(w)\,,
\end{equation}
where $\lambda_{\rm c}>1$ is a tunable stub parameter. \ch{Using} this definition \ch{to compute the second term in} \eqref{eq:PCEOM3}, we get
\begin{subequations}
\label{eq:TTT}
\begin{align}
        \wc\left(T,l_2(T,T)\right)&=\langle c\Bar{c}\mathbb{V}(0,\Bar{0})\vert c_{0}^{-}b_{0}^{-}\delta(L_{0}^{-})\vert f_1\circ  c\Bar{c}\mathbb{V}(0,\Bar{0})f_2\circ  c\Bar{c}\mathbb{V}(0,\Bar{0})\rangle\\[1.8mm]
        &=\langle I_{\rm c}\circ  c\Bar{c}\mathbb{V}(0,\Bar{0}) f_1\circ  c\Bar{c}\mathbb{V}(0,\Bar{0})f_2\circ  c\Bar{c}\mathbb{V}(0,\Bar{0})\rangle_{\mathbb{C}}\\
        &=-\tensor{C}{_{\mathbb{VVV}}}\left(\frac{3}{2}\lambda_{\rm c}\right)^{6y}\,,  
    \end{align}
    \end{subequations}
where, in the first line, we have reabsorbed the projector onto the level-matched Hilbert space in the bra, and, in the second line, we introduced the \ch{closed-string BPZ inverse map} defined as $I_{\rm c}(w)=\frac{1}{w}$. 

\ch{Summarizing our progress up to this point,} plugging back \eqref{eq:TT} and \eqref{eq:TTT} into \eqref{eq:PCEOM3} and expressing $t(y)$ as a power-series expansion, we obtain
\begin{equation}
    y-\frac{1}{2}(t_1 y+t_2 y^2)\left(1+6y\log\left(\frac{3}{2}\lambda_{\rm c}\right)\right)\tensor{C}{_{\mathbb{VVV}}}+\frac{1}{3!}t_{1}^{2}y^2  {\cal A}_{TTTT}+\mathcal{O}(y^3)=0\,.
\end{equation}
\ch{This enables one to determine the coefficients $t_1$ and $t_{2}$ as}
\begin{subequations}
\begin{align}
    t_1&=\frac{2}{\tensor{C}{_{\mathbb{VVV}}}}\,,\label{eq:t1}\\
    t_2&=-\left(\frac{2}{\tensor{C}{_{\mathbb{VVV}}}}\right)^2\left(3\tensor{C}{_{\mathbb{VVV}}}\log\left(\frac{3}{2}\lambda_{\rm c}\right)\right)+\frac{1}{3!}\left(\frac{2}{\tensor{C}{_{\mathbb{VVV}}}}\right)^3{\cal A}_{TTTT}\vert_{y=0}\,.\label{eq:t2}
\end{align}
\end{subequations}
\ch{It remains to evaluate the amplitude ${\cal A}_{TTTT}$ of four slightly relevant operators. This} is one of the main results of \cite{Scheinpflug:2023lfn}. \ch{It turns out that in the large stub limit $\lambda_{\rm c}\to \infty$, one can write it as}
 \begin{equation}
 \label{eq:Atttt}
     {\cal A}_{TTTT}=12\left(\tensor{C}{_{\mathbb{VVV}}}\right)^2\log\left(\frac{3}{2}\lambda_{\rm c}\right)+{\cal A}_{TTTT}^{\rm f.p.}\,,
 \end{equation}
\ch{where ${\cal A}_{TTTT}^{\rm f.p.}$ is a finite number.} Indeed, in computing this amplitude, there are contact \ch{divergences} which can be regularized through a cutoff naturally provided by the closed stub parameter in the limit $\lambda_{\rm c}\to \infty$. The idea is then to expand the four-tachyon fundamental vertex in conformal blocks and then add and subtract divergent terms to obtain a finite part ${\cal A}_{TTTT}^{\rm f.p.}$ in which we can truly perform the limit $\lambda_{\rm c}\to \infty$.
The remaining divergent terms explicitly depend on the closed-string stub parameter $\lambda_\mathrm{c}$. 
Note that by considering the limit $\lambda_\mathrm{c}\to\infty$, we do not lose generality because any observable which we compute must be independent of the off-shell SFT data. \ch{We will explicitly see this at work} in the following \ch{sections, where we will prove} that the classical on-shell disk action does not depend on $\lambda_{\rm c}$. 


Finally, by inserting \eqref{eq:Atttt} into \eqref{eq:t2}, we obtain
 \begin{equation}
 \label{eq:tt2}
     t_2=\left(\frac{{\cal A}^{\rm f.p.}_{TTTT}}{3 \tensor{C}{_{\mathbb{VVV}}}}+ \tensor{C}{_{\mathbb{VVV}}}\log\left(\frac{3}{2}\lambda_{\rm c}\right)\right)\left(\frac{2}{ \tensor{C}{_{\mathbb{VVV}}}}\right)^2\,.
 \end{equation}
\ch{Hence}, using \eqref{eq:splitsol}, \eqref{eq:ansatz}, \eqref{eq:R}, \eqref{eq:t1} and \eqref{eq:tt2}, we find that the classical closed-string solution \ch{expanded} up to quadratic order in $y$ reads 
 \begin{align}
 \label{eq:CSOL}
     \Phi^{\ast}(y)&=\left[\frac{2y}{    \tensor{C}{_{\mathbb{VVV}}} 
  }+\left(\frac{{\cal A}^{\rm f.p.}_{TTTT}}{3 \tensor{C}{_{\mathbb{VVV}}}}+ \tensor{C}{_{\mathbb{VVV}}}\log\left(\frac{3}{2}\lambda_{\rm c}\right)\right)\left(\frac{2y}{ \tensor{C}{_{\mathbb{VVV}}}}\right)^2\right]c\Bar{c}\mathbb{V}+\nonumber\\
  &\hspace{5cm}-\frac{1}{2}\left(\frac{2y}{    \tensor{C}{_{\mathbb{VVV}}} 
  }\right)^2\frac{b_{0}^{+}}{L_{0}^{+}}\bar{P}l_2(c\Bar{c}\mathbb{V},c\Bar{c}\mathbb{V})+\mathcal{O}(y^3)\,.
 \end{align}
\ch{When one substitutes the solution \eqref{eq:CSOL} into the sphere part of the string field theory action, all dependence on the stub parameter $\lambda_\mathrm{c}$ indeed cancels, as it should, since the on-shell value of the action is a gauge-invariant (observable) quantity. In \cite{Scheinpflug:2023lfn}, this quantity has been conjectured to be related to the change in the central charge of the matter CFT under the short bulk RG flow triggered by $\mathbb{V}$, and this has been non-trivially verified up to $\mathcal{O}(y^4)$ in the case of Zamolodchikov flows of large $m$ minimal models.}

\subsection{Fate of probe D-branes under short bulk RG flows}\label{sec:3.3}

{\ch{In this section, we will make a key intermediate step towards computing the disk part of the on-shell action in the situation when the closed-string sector (which is coupled to a probe D-brane system) is undergoing a short RG flow in the matter sector: we will compute the solution $\Psi^\ast(y)$ to the tadpole-sourced open-string equation of motion \eqref{eq:OSEOM}.}} \ch{We will see that since we are interested in expanding the disk action up to second order in $y$, we can truncate the open-string solution at first order in $y$.} \ch{This is because for a critical open-string background, the disk open-string tadpole vanishes, so that second order terms in the open-string solution $\Psi^\ast(y)$ only start contributing into the on-shell disk action at third order in $y$.}

The initial open-string background \ch{can be} defined by a \ch{consistent} boundary CFT. \ch{This in turn can} characterized by a boundary state which can be written as a tensor product of boundary states for each sector of the bulk theory, that is
\begin{equation}
    \|B_{\rm tot}\rangle \!\rangle=\|B_{0}\rangle \!\rangle\otimes \|B_{\rm aux}\rangle \!\rangle\otimes \|B_{\rm gh}\rangle \!\rangle\,.
\end{equation}
\ch{In order to find $\Psi^\ast(y)$, we will follow a strategy similar to the above-described case of the pure closed-string background. Namely,} we will split the open-string field into two orthogonal components through the application of the projector $P$, which projects onto the sub-space \ch{of the open-string Hilbert space} which is spanned by the $L_{0}$ eigenvectors with eigenvalues $0$ and $-y$. \ch{In other words, we write}
\begin{equation}
    \Psi=P\Psi + \bar{P}\Psi\coloneqq w+r\,.
\end{equation}
\ch{For the component $w$ (which lies in the image of $P$)}, we consider an ansatz proportional to the nearly marginal boundary field
\begin{equation}
\label{eq:wansatz}
    w = \tau(y) \,c\,\mathbbm{v}\,.
\end{equation}
\ch{To obtain an equation fixing the function $\tau(y)$,} we \ch{consider} the tadpole-sourced open-string equation of motion evaluated on \ch{the classical closed-string background} $\Phi^{\ast}(y)$ which is given by \eqref{eq:CSOL}. \ch{Assuming that the open-string solution $\Psi^\ast$ will start at order $y$, this can be truncated as}
\begin{align}
       {\rm EOM}_{\Phi^{\ast}(y)}[\Psi]&=Q_{\rm o}\Psi+m_{0,2}(\Psi,\Psi)+\nonumber\\
       &\hspace{2cm}+m_{1,0}(\Phi^{\ast}(y))+m_{1,1}(\Phi^{\ast}(y), \Psi)+m_{2,0}(\Phi^{\ast}(y),\Phi^{\ast}(y))+\mathcal{O}(y^3)\,.
\end{align}
\ch{Solving this equation can be facilitated by considering its projections into the image of $P$ and its orthogonal complement, namely}
\begin{subequations}
\begin{align}
\label{eq:POEOM}
    P\,{\rm EOM}_{\Phi^{\ast}(y)}[\Psi]&=Q_{\rm o}w+Pm_{0,2}(\Psi,\Psi)+\nonumber\\
       &\hspace{1cm}+Pm_{1,0}(\Phi^{\ast}(y))+Pm_{1,1}(\Phi^{\ast}(y), \Psi)+Pm_{2,0}(\Phi^{\ast}(y),\Phi^{\ast}(y))+\mathcal{O}(y^3)\,,\\
    \bar{P}\,{\rm EOM}_{\Phi^{\ast}(y)}[\Psi]&=Q_{\rm o}r+\bar{P}m_{0,2}(\Psi,\Psi)+\nonumber\\
       &\hspace{1cm}+\bar{P}m_{1,0}(\Phi^{\ast}(y))+\bar{P}m_{1,1}(\Phi^{\ast}(y), \Psi)+\bar{P}m_{2,0}(\Phi^{\ast}(y),\Phi^{\ast}(y))+\mathcal{O}(y^3)\,.
\end{align}
\end{subequations}
\ch{Since the open-string BRST operator $Q_\mathrm{o}$ is invertible on the image of $\bar{P}$, the second equation can now be recursively solved in the Siegel gauge $b_0=0$ as}
\begin{equation}
    r^{\ast}(y)=-\frac{b_0}{L_0}\bar{P}m_{1,0}(\Phi^{\ast}(y))+\mathcal{O}(y^2)=-\frac{2y}{\tensor{C}{_{\mathbb{VVV}}}}\frac{b_0}{L_0}\bar{P}m_{1,0}(c\bar{c}\mathbb{V})+\mathcal{O}(y^2)\,.\label{eq:massiveSol}
\end{equation}
\ch{Notice that, in contrast to the closed-string case, the component of the open-string solution lying in the image of $\bar{P}$ now starts at order $y$ and not $y^2$. This is due to the presence of a closed-string sourced open-string tadpole in the disk action.} 

\ch{It remains to solve the equation of motion \eqref{eq:POEOM} for $w$ (and therefore for $\tau(y)$). First, 
note that when the slightly relevant boundary field $\mathbbm{v}$ is not part of the spectrum of the matter BCFT, the image of the projector $P$ can be taken as empty.\footnote{\ch{This is provided that we choose to work with vertices which do not generate the pure-ghost boundary field $\p c$.}} In such cases, the equation of motion \eqref{eq:POEOM} is simply absent and the open-string solution $\Psi^\ast (y)$ lies purely in the image of $\bar{P}$.} 
\ch{On the other hand, when  the boundary field $\mathbbm{v}$ \emph{is} part of the matter BCFT, then, under our assumption \eqref{eq:assumption}, we can see that the $P$-component $w$ of $\Psi^\ast(y)$ will start at $\mathcal{O}(y)$.} \ch{Indeed, the contribution}
\begin{equation}
  \ch{  Q_{\rm o}w=\tau(y) Q_{\rm o} c\,\mathbbm{v}=y \tau(y)  \,c\p c \,\mathbbm{v} = \mathcal{O}(y^2)\,.}
\end{equation}
\ch{to the r.h.s. of \eqref{eq:POEOM} then generally comes at the same order as  $Pm_{0,2}(\Psi^\ast (y),\Psi^\ast (y))$, as well as $Pm_{1,0}(\Phi^\ast (y))$ which acts as a source. Hence, if we put $w(y)=\tau_1 y\, c \,\mathbbm{v}+\mathcal{O}(y^2)$, the equation of motion \eqref{eq:POEOM} can be balanced and, at least in principle, solved for the coefficients $\tau_{i}$. This is how our assumption \eqref{eq:assumption} guarantees perturbativity of the bulk-induced boundary deformation.}

\ch{However, it turns out that in order to achieve our goal of finding the on-shell disk action up to second order in $y$, we do not have to be interested in finding the explicit expression for the solution $w^{\ast}(y)=P\Psi^{\ast}(y)$ because it only would have started to contribute at $\mathcal{O}(y^3)$.} Indeed, the \ch{leading contribution to \eqref{eq:Lam}} involving the \ch{component of the open-string solution lying in the image of $P$ would go as}
\begin{equation}
     \wo\big[w^{\ast}(y),m_{1,0}(\Phi^{\ast}(y))\big]\sim y^2 \frac{\tensor{B}{_{\mathbb{V}\mathbbm{v}}}}{g}+\dots \sim \mathcal{O}(y^3)
\end{equation}
\ch{where two powers of $y$ come from expanding the string fields $w^{\ast}(y)$ and $\Phi^{\ast}(y)$ in $y$, while another power of $y$ comes from the bulk-boundary OPE coefficient $\tensor{B}{_{\mathbb{V}\mathbbm{v}}}/g\sim \mathcal{O}(y)$ as per our assumption \eqref{eq:assumption}.} \ch{On the other hand,} the $\bar{P}$ component of the solution contributes to \eqref{eq:Lam} at order $y^2$
\begin{equation}
   \wo\big[r^{\ast}(y),m_{1,0}(\Phi^{\ast}(y))\big]\sim y^2 \frac{(\tensor{B}{_{\mathbb{V}\mathbbm{1}}})^2}{g}+\dots \sim \mathcal{O}(y^2)\,,
\end{equation}
because generally $\tensor{B}{_{\mathbb{V}\mathbbm{1}}}(y)/g(y)\sim \mathcal{O}(y^0)$.

\ch{To summarize, for the purposes of evaluating the on-shell disk action up to second order in $y$, we may take the open string solution to read
\begin{equation}
\label{eq:OSOL}
    \Psi^{\ast}(y)=-\frac{2y}{\tensor{C}{_{\mathbb{VVV}}}}\frac{b_0}{L_0}\bar{P}m_{1,0}(c\bar{c}\mathbb{V})+\ldots\,,
\end{equation}
where the dots $\ldots$ represent not only $\mathcal{O}(y^2)$ terms, but also terms at order $y$ which, however, would have contributed at cubic order into the on-shell disk action.}

\section{\ch{Disk action of short RG flows}}\label{sec:4}
In this section, we will compute (up to quadratic order in $y$) the disk action \ch{of open-closed SFT} evaluated on the classical solutions $\Phi^{\ast}(y)$ and $\Psi^{\ast}(y)$ \ch{which were constructed in section \ref{sec:3} to describe short RG-flows in the matter (B)CFT.} According to the generalised Sen's conjecture \eqref{eq:GenSenConj}, this will allow us to obtain the induced \ch{change in the ratio of the worldsheet boundary state $g$-function and the string coupling constant in terms of CFT data.}

\subsection{Choice of vertices and initial setup}\label{sec:4.1}
In order to evaluate the on-shell disk action, the first step is to \ch{properly define} the vertices that appear in \eqref{eq:CSOL}, \eqref{eq:OSOL}, as well as on the r.h.s.\ of \eqref{eq:Lam}. Specifically, we will use the definition of multi-string products given in \cite{cosmo}, which is based on the construction of fundamental vertices using ${\rm SL}(2,\mathbb{C})$ maps. The reason why this construction is particularly useful is that it enables us to set the multi-string product $l_{1,0}$ to zero, as demonstrated in \cite{cosmo}. Indeed, in such a \ch{setup}, the interior of the moduli space, associated with the amplitude with two \ch{closed-string} punctures on the disk, is fully covered through Feynman diagrams given by fundamental vertices \ch{arising} at lower order in \ch{string perturbation theory}. In other words, \ch{the ${\rm SL}(2,\mathbb{C})$ products turn out to} satisfy the corresponding homotopy relation without $l_{1,0}$, namely
\begin{equation}
\label{homotopyrel}
    l_2l_{0,0}+l_{0,1}m_{1,0}=0\,.
\end{equation}
Let us then \ch{explicitly} list the string products which appear in \eqref{eq:CSOL}, \eqref{eq:OSOL} and \eqref{eq:Lam}. \ch{These were} defined in \cite{cosmo} as
\begin{subequations}
\begin{align}
\label{l2}
    \vert l_2(\Phi_1,\Phi_2)\rangle &= b_{0}^{-}\delta(L_{0}^{-})f_1\circ \Phi_1(0,\Bar{0})f_2\circ \Phi_2(0,\Bar{0})\vert 0\rangle_{{\rm SL}(2,\mathbb{C})}\,,\\[1.7mm]
    \label{l00}
    \vert l_{0,0}\rangle&=\frac{1}{(2\pi i)^2}\lambda_{\rm b}^{-L_{0}^{+}}\| B_0\rangle\!\rangle\,,\\
    \label{m10}
    (-)^{d(\Phi)}\vert m_{1,0}(\Phi)\rangle&=\frac{1}{2\pi i} \widetilde{\left[m\circ \Phi(0,\Bar{0}) \right]} \vert 0\rangle_{{\rm SL}(2,\ch{\mathbb{R}})}\,,
\end{align}
\end{subequations}
where $\lambda_{\rm b}>1$ is a tunable stub parameter and $\widetilde{(\cdots)}$ means that the \ch{bulk-boundary OPE} is understood \ch{to have been performed}. \ch{Finally, the} ${\rm SL}(2,\mathbb{C})$ maps \ch{$f_1(w)$, $f_2(w)$ and $m(w)$ are defined as} 
\begin{subequations}
\begin{align}
\label{f}
    f_1(w)&=\frac{1}{\lambda_{\rm c}}\frac{w-\lambda_{\rm c}}{3\lambda_{\rm c}+w}=-f_2(w)\,,\\
    \label{m}
    m(w)&=\frac{i}{\lambda_{\rm o}}\frac{1+\frac{w}{\beta_2}}{1+\frac{w}{\beta_1}}\,,
\end{align}
\end{subequations}
in which $\lambda_{\rm c}>1$ is a free \ch{stub} parameter whereas $\lambda_{\rm o}$, $\beta_1$ and $\beta_2$ are fixed by the homotopy relation \eqref{homotopyrel}  as 
\begin{subequations}
\begin{align} 
    \lambda_{\rm o}&=\frac{3\lambda_{\rm b}\lambda_{\rm c}+1}{3\lambda_{\rm b}\lambda_{\rm c}-1}\,,\label{ocinterpolation}\\
    \beta_1&=\frac{3\lambda_{\rm b}\lambda_{\rm c}+1}{3\lambda_{\rm b}\lambda_{\rm c}-1}\lambda_{\rm c}\,,\\
    \beta_2&=\frac{3\lambda_{\rm b}\lambda_{\rm c}-1}{3\lambda_{\rm b}\lambda_{\rm c}+1}\lambda_{\rm c}\,.
\end{align}
\end{subequations}
Notice that \eqref{ocinterpolation} gives a relation between the \ch{open-string} and \ch{closed-string} stub parameters, \ch{which is explicitly showing how the open-closed SFT naturally interpolates between the two limits where the dynamics is mostly dominated by the open and closed strings, respectively \cite{Zwiebach:1992bw,Firat:2023gfn}.} Indeed, the large \ch{closed-string} stub limit $\lambda_{\rm c}\to \infty$ implies the short \ch{open-string} stub limit $\lambda_{\rm o}\to \ch{\mathcal{O}(1)}$, which means that the moduli space will be mostly covered by open-string propagators and closed-string fundamental \ch{vertex regions}. Conversely, if $\lambda_{\rm o}\to \infty$, then $\lambda_{\rm c}\to \ch{\mathcal{O}(1)}$, so that the moduli space is mostly covered by \ch{closed-string} propagators and \ch{open-string} fundamental vertices. As we have advertised in the previous sections, we will focus on the former limit.

Given this particular choice of SFT vertices, we can further simplify the expression \eqref{eq:Lam} by setting $l_{1,0}=0$, namely
\begin{equation}
    \frac{1}{2\pi^2}\bigg[\frac{g^\ast}{g}\bigg(\frac{g_\mathrm{s}^\ast}{g_\mathrm{s}}\bigg)^{-1}-1\bigg]= \frac{1}{g}\,\wc\left(\Phi^{\ast}(y),l_{0,0}\right)+\frac{1}{2}\frac{1}{g}\,\wo\left(\Psi^{\ast}(y),m_{1,0}(\Phi^{\ast}(y))\right)+\mathcal{O}(y^3)\,.\label{eq:OnShellDisk0}
\end{equation}
\ch{Substituting the explicit expressions \eqref{eq:CSOL} and \eqref{eq:OSOL} for the classical solutions $\Phi^{\ast}(y)$ and $\Psi^{\ast}(y)$ into the r.h.s.\ of \eqref{eq:OnShellDisk0} and carefully collecting all contributions, we obtain}
\begin{align}
\label{diskaction}
    &\frac{1}{2\pi^2}\bigg[\frac{g^\ast}{g}\bigg(\frac{g_\mathrm{s}^\ast}{g_\mathrm{s}}\bigg)^{-1}-1\bigg]=\nonumber\\
    &\hspace{1cm}=\left[\frac{2y}{    \tensor{C}{_{\mathbb{VVV}}} 
  }+\left(\frac{{\cal A}^{\rm f.p.}_{TTTT}}{3 \tensor{C}{_{\mathbb{VVV}}}}+ \tensor{C}{_{\mathbb{VVV}}}\log\left(\frac{3}{2}\lambda_{\rm c}\right)\right)\left(\frac{2y}{ \tensor{C}{_{\mathbb{VVV}}}}\right)^2\right] {\cal A}^{\rm disk}_{T}-\frac{1}{2} \left(\frac{2y}{ \tensor{C}{_{\mathbb{VVV}}}}\right)^2 {\cal A}^{\rm disk}_{TT}+O(y^3)\,,  
\end{align}
where ${\cal A}^{\rm disk}_{T}$ and ${\cal A}^{\rm disk}_{TT}$ are the \ch{zero-momentum one- and two-point amplitudes of the nearly on-shell tachyon $c\bar{c}\mathbb{V}$} on the disk, \ch{which are rescaled} by the ratio of the initial $g$-function and the string coupling constant. \ch{That is to say, we define}
\begin{subequations}
\begin{align}
    {\cal A}^{\rm disk}_{T}&\coloneqq\frac{1}{g}\,\wc\left(c\Bar{c}\mathbb{V}, l_{0,0}\right)\,,\label{At}\\
  {\cal A}^{\rm disk}_{TT}&\coloneqq \frac{1}{g}\,\wc\left(\frac{b_{0}^{+}}{L_{0}^{+}}\Bar{P}l_2( c\Bar{c}\mathbb{V}, c\Bar{c}\mathbb{V}),l_{0,0}\right)+\frac{1}{g}\,\wo\left(\frac{b_0}{L_0}\Bar{P}m_{1,0}( c\Bar{c}\mathbb{V}),m_{1,0}( c\Bar{c}\mathbb{V})\right)\,. \label{Att}
\end{align}
\end{subequations}
Looking at \eqref{diskaction}, it \ch{would naively appear} that the on-shell disk action \ch{could} depend on the stub parameter $\lambda_{\rm c}$. \ch{This would stand in contradiction with the interpretation of the on-shell disk action as an observable quantity because as such, it should not depend on the choice of the local coordinate maps.}
Therefore, it \ch{will be crucial to prove that, after careful evaluation of the two amplitudes ${\cal A}^{\rm disk}_{T}$ and ${\cal A}^{\rm disk}_{TT}$, all dependence on $\lambda_\mathrm{c}$ drops out, so that one ends up with an expression that is independent of the off-shell SFT data. In the following subsection, we will explicitly compute these two amplitudes in the large closed-string stub limit $\lambda_\mathrm{c}\to \infty$.}

\subsection{Computing  \texorpdfstring{$ {\cal A}^{\rm disk}_{T}$}{TEXT} and  \texorpdfstring{$ {\cal A}^{\rm disk}_{TT}$}{TEXT}}\label{sec:4.2}
Let us begin by calculating the amplitude ${\cal A}^{\rm disk}_{T}$ of one tachyon on the disk. By \ch{substituting into \eqref{At}} the explicit expression \eqref{l00} for the closed-string tadpole and rewriting the symplectic form as the BPZ inner product, we obtain
\begin{equation}
\label{At1}
     {\cal A}^{\rm disk}_{T}=-\frac{1}{4\pi^2 g}\langle  c\Bar{c}\mathbb{V}(0,\Bar{0})\vert c_{0}^{-}\lambda_{\rm b}^{-L_{0}^{+}}\| B_{0}\rangle\!\rangle 
     =\frac{1}{4\pi^2}\frac{\tensor{B}{_{\mathbb{V}\mathbbm{1}}}}{g}\lambda_{\rm b}^{2y},
\end{equation}
where in the second equality, we acted with the BPZ-even operator $\lambda_{\rm b}^{L_{0}^{+}}$ on the bra and we also computed the ghost correlator
\begin{equation}
    \langle c\bar{c}(0,\bar{0})\vert c_{0}^{-}\| B_{\mathrm{ghost}}\rangle\!\rangle=\frac{1}{8}\big\langle c(i)c(-i)\left(\p c(i)-\p c(-i)\right) \big\rangle_{\rm UHP}=-1\,.
\end{equation}
As far as the matter contribution is concerned, we \ch{only have to deal with} the one-point function on the disk $\langle \mathbb{V}(0,\Bar{0})\| B_{0}\rangle\!\rangle$, which is \ch{identically} equal to $\tensor{B}{_{\mathbb{V}\mathbbm{1}}}$. \ch{Finally, expanding the stub contribution up to first order in $y$, we end up with the expression}
\begin{equation}
\label{At2}
    {\cal A}^{\rm disk}_{T}=\frac{1}{4 \pi^2}\left[\frac{\tensor{B}{_{\mathbb{V}\mathbbm{1}}}}{g}+\left(\frac{1}{2}\frac{\tensor{B}{_{\mathbb{V}\mathbbm{1}}}}{g} \tensor{C}{_{\mathbb{VVV}}}\log\left(\lambda_{\rm b}\right)\right)\left(\frac{2y}{ \tensor{C}{_{\mathbb{VVV}}}}\right)\right]+\mathcal{O}(y^2)\,,
\end{equation}
\ch{which is exact up to linear order in $y$. This precision is sufficient as we recall that $\mathcal{A}_{T}^{\mathrm{disk}}$ is already multiplied by $y$ in the expression \eqref{diskaction} for the on-shell disk action.}

\ch{Second, let us} deal with \ch{the amplitude} ${\cal A}^{\rm disk}_{TT}$ \ch{which was defined in} \eqref{Att}. In this case, we will only be interested in the leading order $\mathcal{O}(y^0)$ because in  \eqref{diskaction}, ${\cal A}^{\rm disk}_{TT}$ appears already multiplied by $y^2$. \ch{Focusing first on the closed-string exchange, we can write}
\begin{subequations}
\label{closedchannel0}
\begin{align}
     &\frac{1}{g}\,\wc\left(\frac{b_{0}^{+}}{L_{0}^{+}}\Bar{P}l_2( c\Bar{c}\mathbb{V}, c\Bar{c}\mathbb{V}),l_{0,0}\right)= \nonumber\\
     &\hspace{2cm} =\frac{1}{g}\langle l_{0,0}\vert c_{0}^{-}\frac{b_{0}^{+}}{L_{0}^{+}}\Bar{P}\vert l_2( c\Bar{c}\mathbb{V}(0,\Bar{0}), c\Bar{c}\mathbb{V}(0,\Bar{0}))\rangle  \\
     &\hspace{2cm} =-\frac{1}{4 \pi^2 g}\langle\!\langle B_0\| c_{0}^{-}\lambda_{\rm b}^{-L_{0}^{+}}\frac{b_{0}^{+}}{L_{0}^{+}}\Bar{P}b_{0}^{-}c_{0}^{-}\vert l_2( c\Bar{c}\mathbb{V}(0,\Bar{0}), c\Bar{c}\mathbb{V}(0,\Bar{0}))\rangle \\
     &\hspace{2cm} =-\frac{1}{4 \pi^2 g}\langle\!\langle B_0\| c_{0}^{-}\lambda_{\rm b}^{-L_{0}^{+}}\vert \xi_{i}\rangle\langle \xi_{i}^{c}\vert\frac{b_{0}^{+}}{L_{0}^{+}}\Bar{P}b_{0}^{-}\vert \xi_{j}^{c}\rangle\langle \xi_{j}\vert c_{0}^{-}\vert l_2( c\Bar{c}\mathbb{V}(0,\Bar{0}), c\Bar{c}\mathbb{V}(0,\Bar{0}))\rangle\,.
\end{align}
\end{subequations}
Here, in the second equality, we wrote out the closed-string tadpole in its explicit form \eqref{l00} and we \ch{also made manifest the projection onto the $b_0^{-}$ exact states in the intermediate channel}. Finally, in the last equality, we made two insertions of the identity operator $1_{{\cal H}_{\rm c}}=\vert \xi_i\rangle\langle \xi_{i}^{c}\vert=\vert \xi_{i}^{c} \rangle\langle \xi_i\vert$ on the Hilbert space ${\cal H}_{\rm c}$ of the level-matched closed-string states (where sums over repeated indices are understood). \ch{The basis vectors $|\xi_i\rangle$ and their BPZ duals $|\xi_i^{c}\rangle$} are normalized so that $\langle \xi_{i}^{c}\vert\xi_{j}\rangle=\langle \xi_{i}\vert\xi_{j}^{c}\rangle=\delta_{ij}$. 

Following the argument described in \cite{Scheinpflug:2023lfn}, \ch{thanks to the projector $\bar{P}$ in front of the closed-string propagator, the only contribution received by \eqref{closedchannel0} in the large stub limit comes from the propagation of the \ch{universal tachyon} $\vert \xi_j\rangle=\vert \xi_i\rangle=\vert c_{1}\Bar{c}_{1}\rangle$.} \ch{Hence, the part \eqref{closedchannel0} of the amplitude ${\cal A}^{\rm disk}_{TT}$ associated with the closed-string exchange can be rewritten as}
\begin{align}
\label{closedchannel1}
        &\frac{1}{g}\,\wc \left(\frac{b_{0}^{+}}{L_{0}^{+}}\Bar{P}l_2\left( c\Bar{c}\mathbb{V}, c\Bar{c}\mathbb{V}\right),l_{0,0}\right)=\nonumber \\
     &\hspace{0.4cm}=-\frac{1}{4 \pi^2 g}\langle\!\langle B_0\| c_{0}^{-}\lambda_{\rm b}^{-L_{0}^{+}}\vert c_{1}\Bar{c}_{1}\rangle\langle c_{0}\Bar{c}_{0}c_{1}\Bar{c}_{1}\vert\frac{b_{0}^{+}}{L_{0}^{+}}\Bar{P}b_{0}^{-}\vert c_{0}\Bar{c}_{0}c_{1}\Bar{c}_{1}\rangle\langle c_{1}\Bar{c}_{1}\vert c_{0}^{-}\vert l_2( c\Bar{c}\mathbb{V}(0,\Bar{0}), c\Bar{c}\mathbb{V}(0,\Bar{0}))\rangle\,.
\end{align}
\ch{Let us now separately evaluate the three factors on the r.h.s.\ of \eqref{closedchannel1}.} The computation of the first one is \ch{entirely} analogous to what was done in \eqref{At1} \ch{provided that we replace $\mathbb{V}$ with the identity in the matter sector. We then obtain}
\begin{equation}
    -\frac{1}{4 \pi^2 g}\,\langle\!\langle B_0\| c_{0}^{-}\lambda_{\rm b}^{-L_{0}^{+}}\vert c_{1}\Bar{c}_{1}\rangle=\frac{1}{4 \pi^2 }\,\lambda_{\rm b}^{2}\,.
\end{equation}
Regarding the second \ch{factor}, we trivially obtain
\begin{equation}
    \langle c_{0}\Bar{c}_{0}c_{1}\Bar{c}_{1}\vert\frac{b_{0}^{+}}{L_{0}^{+}}\Bar{P}b_{0}^{-}\vert c_{0}\Bar{c}_{0}c_{1}\Bar{c}_{1}\rangle=1\,,
\end{equation}
\ch{while} the third \ch{factor} can be computed using \eqref{l2} and \eqref{f} as
\begin{subequations}
    \begin{align}
  \langle c_{1}\Bar{c}_{1}\vert c_{0}^{-}\vert l_2( c\Bar{c}\mathbb{V}(0,\Bar{0}), c\Bar{c}\mathbb{V}(0,\Bar{0}))\rangle&=\langle c_{1}\Bar{c}_{1}\vert c_{0}^{-}b_{0}^{-}\delta(L_{0}^{-})\vert f_1\circ  c\Bar{c}\mathbb{V}(0,\Bar{0})f_2\circ  c\Bar{c}\mathbb{V}(0,\Bar{0})\rangle   \\[2mm]
  &=\big\langle I_{\rm c}\circ  c\Bar{c}(0,\Bar{0})\, f_1\circ  c\Bar{c}\mathbb{V}(0,\Bar{0})\, f_2\circ  c\Bar{c}\mathbb{V}(0,\Bar{0})\big\rangle_{\mathbb{C}}\\
  &=-\left(\frac{3}{2}\lambda_{\rm c}\right)^{2}+\mathcal{O}(y)\,,
    \end{align}
\end{subequations}
where $I_{\rm c}(w)=\frac{1}{w}$ \ch{denotes the closed-string BPZ inversion}. Putting all these results together, we get
\begin{equation}
    \label{closedchannel}
      \frac{1}{g}\,\wc\left(\frac{b_{0}^{+}}{L_{0}^{+}}\Bar{P}l_2( c\Bar{c}\mathbb{V}, c\Bar{c}\mathbb{V}),l_{0,0}\right)=-\frac{1}{4\pi^2 }\left(\frac{3}{2}\lambda_{\rm c}\lambda_{\rm b}\right)^{2}+\mathcal{O}(y)
\end{equation}
\ch{for the closed-string channel.}

\ch{Let us now focus on computing the channel of $ {\cal A}^{\rm disk}_{TT}$ which is associated with the open-string exchange (second term in \eqref{Att}). Substituting the explicit form \eqref{m10} of the open-closed product $m_{1,0}$, we can first express}
    \begin{align}
    \label{openchannel0}
      &\frac{1}{g}\,\wo\left(\frac{b_0}{L_0}\Bar{P}m_{1,0}\left( c\Bar{c}\mathbb{V}\right),m_{1,0}\left( c\Bar{c}\mathbb{V}\right)\right)=\nonumber \\
      &\hspace{3cm}= -\frac{1}{4 \pi^2 g} \int_{0}^{1}\frac{dt}{t}t^{\epsilon_o}\big\langle I_{\rm o}\circ m \circ  c\Bar{c}\mathbb{V}(0,\Bar{0}) \,b_{0}\,t^{L_{0}}\, m \circ  c\Bar{c}\mathbb{V}(0,\Bar{0})\big\rangle_{\rm UHP}\biggr\vert_{\epsilon_{o}\to 0}\,,
    \end{align}
where we wrote the Siegel propagator in the Schwinger representation and \ch{introduced the open-string BPZ inversion $I_{\text{o}}(w)=-\frac{1}{w}$}. Furthermore, we introduced \ch{the factor} $t^{\epsilon_{o}}$ to \ch{regulate the divergences which generally arise near the open-string degeneration $t\to 0$ due to propagation of open-string tachyons. In our particular case, the only divergence will come} from the identity channel in the bulk-boundary OPE of the matter field $\mathbb{V}$. \ch{Also notice that on the r.h.s.\ of \eqref{openchannel0}, we have dropped the projector $\bar{P}$. This is possible by realizing that
our assumption \eqref{eq:assumption} guarantees that at the order $\mathcal{O}(y^0)$, no states in the image of $P$ propagate in the open-string channel of the amplitude $ {\cal A}^{\rm disk}_{TT}$ so that the $\bar{P}$ insertion in front of the open-string propagator on the l.h.s. of \eqref{openchannel0} becomes indistinguishable from the identity. If \eqref{eq:assumption} were not assumed, the states in the image of $P$ propagating in the open-string channel would have to be removed by hand.
}
\ch{With these provisions in mind, we can implement the action of the conformal maps \eqref{m} on the r.h.s.\ of \eqref{openchannel0} and compute the ghost-part of the correlator.} Keeping only the contributions at the leading order in $y$, we eventually get
\begin{equation}
    \label{openchannel1}
      \frac{1}{g}\,\wo\left(\frac{b_0}{L_0}\Bar{P}m_{1,0}\left( c\Bar{c}\mathbb{V}\right),m_{1,0}\left( c\Bar{c}\mathbb{V}\right)\right) 
       =-\frac{1}{4 \pi^2 g} \int_{0}^{\frac{1}{\lambda_{\rm o}^{2}}}ds\,s^{\epsilon_o}\, 4(s^2-1)\big\langle  \mathbb{V}(i,\Bar{i}) \,\mathbb{V}(is,\Bar{is})\big\rangle_{\rm UHP}\biggr\vert_{\epsilon_{o}\to 0}\,,
\end{equation}
\ch{where the details can be checked by following an analogous computation which was presented in \cite{cosmo}.}


\subsubsection{Open and closed degenerations in \texorpdfstring{$ {\cal A}^{\rm disk}_{TT}$}{TEXT}}\label{sec:4.3}

\ch{At this point, we will focus on making explicit the terms on the r.h.s.\ of \eqref{openchannel1} which diverge as we remove the regulators $\lambda_{\mathrm{o}}$ and $\epsilon_{\mathrm{o}}$, putting particular emphasis on the regime of closed-string degeneration. This regime can be achieved as the Schwinger parameter $s$ approaches 1 and therefore becomes available as we send $\lambda_\mathrm{o}\to 1$. As showcased by the relation \eqref{ocinterpolation}, this is equivalent to the limit $\lambda_\mathrm{b}\lambda_\mathrm{c}\to \infty$ in which the closed-string stub parameters become large. By subtracting and adding the divergent contributions, we will be able to isolate the explicit dependence of \eqref{openchannel1} on the stub parameters $\lambda_\mathrm{b}$, $\lambda_\mathrm{c}$. This will eventually facilitate the proof that the value of the on-shell disk action does not depend on the SFT data.
}


\ch{Let us start by discussing the open-string degeneration $s\to 0$. In this limit, the operator insertion $\mathbb{V}(is,-is)$ in \eqref{openchannel1}
can be replaced by the corresponding bulk-boundary OPE, namely
}
\begin{equation}
    \mathbb{V}(is,-is)\stackrel{s\to 0}{\sim}\frac{1}{(2s)^{2}} \frac{\tensor{B}{_{\mathbb{V}\mathbbm{1}}}}{g}+\frac{1}{2s}\frac{\tensor{B}{_{\mathbb{V}\mathbbm{v}}}}{g}\mathbbm{v}(0)+{\rm reg}\,.
\end{equation}
\ch{Notice that the second term of the above relation would have been annihilated by the $\bar{P}$ appearing in front of the Siegel-gauge propagator on the l.h.s.\ of \eqref{openchannel0} and as such, it would have to be removed by hand when evaluating the open-string channel of the amplitude. However, as per our assumption \eqref{eq:assumption}, we have $\frac{\tensor{B}{_{\mathbb{V}\mathbbm{v}}}}{g}\sim \mathcal{O}(y)$, so that at the order $\mathcal{O}(y^0)$, this is taken care of automatically. As a result, the only divergence arising is the one due to the identity channel, which can be regulated through the parameter $\epsilon_{o}$ in \eqref{openchannel1}. Consequently, in the limit $s\to 0$, one can isolate the divergent contribution to the correlator as}
\begin{equation}
\label{opendiv}
    \big\langle  \mathbb{V}(i,\bar{i}) \mathbb{V}(is,\bar{is})\big\rangle_{\rm UHP}\stackrel{s\to 0}{\sim}\frac{1}{(2s)^{2}} \frac{\tensor{B}{_{\mathbb{V}\mathbbm{1}}}}{g}\big\langle  \mathbb{V}(i,\Bar{i}) \big\rangle_{\rm UHP}=\frac{1}{(4s)^2}\frac{\left(\tensor{B}{_{\mathbb{V}\mathbbm{1}}}\right)^{2}}{g}\,.
\end{equation}
\ch{For the sake of later convenience, let us also introduce an explicit notation}
\begin{equation}
\label{divo}
    {\rm div}_{\rm o}(s)\coloneqq \frac{1}{(4s)^2}\frac{\left(\tensor{B}{_{\mathbb{V}\mathbbm{1}}}\right)^{2}}{g}
\end{equation}
\ch{for the open-string divergence in the 2-point correlator of $\mathbb{V}$. }


The closed string degeneration occurs for $s \to 1$, and it is \ch{therefore regulated} through the stub parameter $\lambda_{\rm o}^{-2}$. \ch{This regulator can be lifted by taking the limit $\lambda_{\rm c} \to \infty$ which, according to \eqref{ocinterpolation}, corresponds to $\lambda_{\rm o} \to 1$.} As \ch{the Schwinger parameter $s$ approaches 1}, we observe a collision between the two bulk fields and, therefore, we can read off the divergent contribution of the correlator to the amplitude $ {\cal A}^{\rm disk}_{TT}$ by using the bulk OPE
\begin{equation}
    \big\langle\mathbb{V}(i,\bar{i})\,\mathbb{V}(is,\bar{is})\big\rangle_{\rm UHP}\stackrel{s\to 1}{\sim} \frac{1}{(1-s)^{4}}\,g+ \frac{1}{4(1-s)^{2}}\,\tensor{C}{_{\mathbb{VVV}}}\tensor{B}{_{\mathbb{V}\mathbbm{1}}}\,.
\end{equation}
\ch{Note that here we are ignoring the simple pole in $1-s$ because thanks to the $c$-ghost measure $4(s^2-1)$, this gives rise to a finite contribution to the amplitude.}
\ch{In an exact analogy with the open-string channel,} we also introduce the notation
\begin{equation}
\label{divc}
    {\rm div}_{\rm c}(s)\coloneqq  \frac{1}{(1-s)^{4}}g+ \frac{1}{4(1-s)^{2}}\tensor{C}{_{\mathbb{VVV}}}\tensor{B}{_{\mathbb{V}\mathbbm{1}}}
\end{equation}
for the closed-string divergence.

Let us then subtract and add \eqref{divo} and \eqref{divc} to the 2-point correlator of $\mathbb{V}$ when evaluating the integral over $s$ in \eqref{openchannel1}. This gives
\begin{align}
\label{openchannel3}
       &\frac{1}{g}\,\wo \left(\frac{b_0}{L_0}\Bar{P}m_{1,0}\left( c\Bar{c}\mathbb{V}\right),m_{1,0}\left( c\Bar{c}\mathbb{V}\right)\right) =\nonumber\\
       &\hspace{2cm}=-\frac{1}{4 \pi^2 }{\cal A}^{\rm f.p.}_{TT} -\frac{1}{4 \pi^2 g} \int_{0}^{\frac{1}{\lambda_{\rm o}^{2}}}ds\,s^{\epsilon_o} 4(s^2-1)\left({\rm div}_{\rm o}(s)+{\rm div}_{\rm c}(s)\right)\biggr\vert_{\epsilon_{o}\to 0}\,,
\end{align}
where we define {\ch a finite contribution}
\begin{equation}
    {\cal A}^{\rm f.p.}_{TT}\coloneqq\frac{1}{g} \int_{0}^{1}ds\, 4(s^2-1)\Big[\big\langle  \mathbb{V}(i,\Bar{i}) \mathbb{V}\left(is,\Bar{is}\right)\big\rangle_{\rm UHP}-{\rm div}_{\rm o}(s)-{\rm div}_{\rm c}(s)\Big]\,,\label{eq:ATTfp}
\end{equation}
\ch{to the amplitude $ {\cal A}^{\rm disk}_{TT}$, in which all regulators can be removed as we are subtracting all sources of divergent behaviour. The value of ${\cal A}^{\rm f.p.}_{TT}$ will of course depend on the details of the particular 2d CFT in which the short RG flow is considered.
On the other hand, the second term in
\eqref{openchannel3} can be evaluated as}
\begin{subequations}
\begin{align}
       & -\frac{1}{4 \pi^2 g} \int_{0}^{\frac{1}{\lambda_{\rm o}^{2}}}ds\,s^{\epsilon_o} 4(s^2-1)\big[{\rm div}_{\rm o}(s)+{\rm div}_{\rm c}(s)\big]\biggr\vert_{\epsilon_{o}\to 0}=\nonumber\\[1mm]
       &\hspace{0.4cm}=-\frac{1}{4 \pi^2 g} \int_{0}^{\frac{1}{\lambda_{\rm o}^{2}}}ds\,s^{\epsilon_o} 4(s^2-1)\left[\frac{1}{(4s)^2}\frac{\left(\tensor{B}{_{\mathbb{V}\mathbbm{1}}}\right)^{2}}{g}+\frac{1}{(1-s)^{4}}g+ \frac{1}{4(1-s)^{2}}\tensor{C}{_{\mathbb{VVV}}}\tensor{B}{_{\mathbb{V}\mathbbm{1}}}\right]\biggr\vert_{\epsilon_{o}\to 0}\\
       &\hspace{0.4cm}=-\frac{1}{4\pi^2}\left[\frac{1}{2}\left(\frac{\tensor{B}{_{\mathbb{V}\mathbbm{1}}}}{g}\right)^2+\frac{1}{2}+\tensor{C}{_{\mathbb{VVV}}}\frac{\tensor{B}{_{\mathbb{V}\mathbbm{1}}}}{g}\left(1+2\log(2)-2\log\left(\frac{3}{2}\lambda_{\rm b}\lambda_{\rm c}\right)\right)-\left(\frac{3}{2}\lambda_{\rm b}\lambda_{\rm c}\right)^2\right]\,.\label{eq:DivPart2}
\end{align}
\end{subequations}
Notice that the last term \ch{in the square brackets of \eqref{eq:DivPart2}} exactly cancels out the closed-string channel contribution \eqref{closedchannel}. 

In total, the amplitude ${\cal A}^{\rm disk}_{TT}$ of two nearly on-shell tachyons on the disk at the order $\mathcal{O}(y^{0})$ becomes
\begin{equation}
\label{Att2}
    {\cal A}^{\rm disk}_{TT}=-\frac{1}{4 \pi^2 }{\cal A}^{\rm f.p.}_{TT}-\frac{1}{4\pi^2}\left[\frac{1}{2}\left(\frac{\tensor{B}{_{\mathbb{V}\mathbbm{1}}}}{g}\right)^2+\frac{1}{2}+\tensor{C}{_{\mathbb{VVV}}}\frac{\tensor{B}{_{\mathbb{V}\mathbbm{1}}}}{g}\left(1+2\log(2)-2\log\left(\frac{3}{2}\lambda_{\rm b}\lambda_{\rm c}\right)\right)\right]\,.
\end{equation}
We observe that all remaining dependence of ${\cal A}^{\rm disk}_{TT}$ on the off-shell SFT data is encapsulated in the logarithmic term
\begin{align}
 \frac{1}{2\pi^2} \,  \tensor{C}{_{\mathbb{VVV}}}\frac{\tensor{B}{_{\mathbb{V}\mathbbm{1}}}}{g}\log\left(\frac{3}{2}\lambda_{\rm b}\lambda_{\rm c}\right)\,.\label{eq:ATTstub}
\end{align}

\subsubsection{The final formula for the \ch{on-shell disk action}}
\label{sec:4.4}
\ch{At this point, we have in place all the quantities needed for writing down the on-shell disk action in terms of CFT data. By substituting the results \eqref{At2} and \eqref{Att2} for the amplitudes ${\cal A}^{\rm disk}_{T}$ and ${\cal A}^{\rm disk}_{TT}$
into the expression \eqref{diskaction} for the disk action, 
we can explicitly see that the stub-dependent part \eqref{eq:ATTstub} of the amplitude ${\cal A}^{\rm disk}_{TT}$ exactly cancels with the $\log(\frac{3}{2} \lambda_{\rm c})$ term in the prefactor of ${\cal A}^{\rm disk}_{T}$ in \eqref{diskaction} coming from the closed-string solution \eqref{eq:CSOL}, as well as with the $\log(\lambda_{\rm b})$ term in the result \eqref{At2} for ${\cal A}^{\rm disk}_{T}$.}
\ch{We therefore obtain the following expression for the on-shell open-closed SFT disk action}
\begin{equation}
\label{eq:finalformula}
    \begin{split}
    &\frac{g^\ast}{g}\bigg(\frac{g_\mathrm{s}^\ast}{g_\mathrm{s}}\bigg)^{-1}=1+ \frac{\tensor{B}{_{\mathbb{V}\mathbbm{1}}}}{g}\left(\frac{y}{    \tensor{C}{_{\mathbb{VVV}}} 
  }\right)+
  \left(\frac{\tensor{B}{_{\mathbb{V}\mathbbm{1}}}}{g}\frac{2{\cal A}^{\rm f.p.}_{TTTT}}{3 \tensor{C}{_{\mathbb{VVV}}}}+\Tilde{{\cal A}}^{\rm f.p.}_{TT}
  \right)\left(\frac{y}{    \tensor{C}{_{\mathbb{VVV}}} 
  }\right)^2+\mathcal{O}(y^3)\,, 
    \end{split}
\end{equation}
\ch{where the shifted finite part $\Tilde{{\cal A}}^{\rm f.p.}_{TT}$ is defined as}
\begin{equation}
    \Tilde{{\cal A}}^{\rm f.p.}_{TT}\coloneqq {\cal A}^{\rm f.p.}_{TT}+\frac{1}{2}\left(\frac{\tensor{B}{_{\mathbb{V}\mathbbm{1}}}}{g}\right)^2+\frac{1}{2}+\tensor{C}{_{\mathbb{VVV}}}\frac{\tensor{B}{_{\mathbb{V}\mathbbm{1}}}}{g}\left(1+2\log(2)\right)\,.\label{atildett}
\end{equation}
\ch{This result is manifestly independent of any off-shell SFT data and is expressed purely in terms of 2d CFT quantities.}


\section{Example: Virasoro minimal models}\label{sec:5}
\ch{In this section, we will present an application of the above-described construction of open-closed SFT solutions
to the explicit example of Virasoro minimal models at $c$ close to 1. From the pure 2d CFT point of view, this setup has already been famously analyzed in the case of pure-bulk RG flows by Zamolodchikov in \cite{Zamolodchikov:1987ti} and, in the case of pure-boundary perturbations, by Recknagel, Roggenkamp and Schomerus in \cite{Recknagel:2000ri}. Here, our aim will be to showcase how open-closed string field theory can be conveniently used to evaluate the response of conformal boundaries to the Zamolodchikov perturbation in the bulk, to the order which would correspond to a \emph{two-loop} calculation in conformal perturbation theory. Apart from confirming the conjectured flows of \cite{Fredenhagen:2009tn} at subleading order in perturbation theory, we will also show that the classical SFT solution describing the Zamolodchikov bulk RG flow changes the string coupling constant at a rate which is given by the same formula (in terms of the sphere 2-point function of the deforming operator) as in the case of the Narain marginal deformations discussed in section \ref{sec:2}.}

\subsection{\ch{Review of bulk and boundary data}}\label{sec:5.1}
\ch{As we have advertised in the introductory paragraph, in this section we will assume that the matter sector of the worldhseet CFT contains a factor given by a unitary A-series Virasoro minimal model. These are 2d CFTs arising at a central charge $0<c<1$ which can be discretely parametrized as}
\begin{equation}
    \label{eq:cm}
    c_{m}=1-\frac{6}{m(m+1)}\,,
\end{equation}
\ch{where $m=3,4,5,\ldots$ For such values of $c$,} the allowed highest weight representations of the Virasoro algebra \ch{can, in turn, be parametrized} by the \ch{Kac labels} $(r,s)$ where $1\le r \le m-1$ and $1\le s \le m$, with the identification $(r,s)\sim (m-r,m+1-s)$. The corresponding weights $h_{(r,s)}$ are given as
\begin{align}
    h_{(r,s)}=\frac{\left[(m+1)r-ms\right]^2-1}{4m(m+1)}\,.\label{eq:hMM}
\end{align}
\ch{As we take $m\to \infty$, the leading contribution to the weights is determined purely by $|r-s|$, as we can write $h_{(r,s)} = \frac{1}{4}(r-s)^2+\mathcal{O}(\frac{1}{m})$.}

\ch{In the bulk, the highest-weight representations $(r,s)$ give rise to the (diagonal) primary fields $\phi_{(r,s)}$ with scaling dimensions $\Delta_{(r,s)}=2h_{(r,s)}= \frac{1}{2}(r-s)^2+\mathcal{O}(\frac{1}{m})$.} 
In particular, to implement a short RG flow in the bulk, \ch{one can} study the \ch{relevant} perturbation induced by \ch{the bulk field} $\mathbb{V}=\phi_{(1,3)}$. Indeed, since its scaling dimension $\Delta_{(1,3)}$ reads 
\begin{equation}
    \Delta_{(1,3)}=2-\frac{4}{m+1}\coloneqq 2(1-y)\,, 
\end{equation}
\ch{we can see that $\phi_{(1,3)}$ approaches marginality as $m\to \infty$ (or, equivalently, as $c_m\to 1$). The expansion parameter $y$ of the corresponding short RG flow (which measures the failure of the relevant perturbing operator $\mathbb{V}$ to be marginal) can thus be identified as}
\begin{equation}
\label{eq:ym}
    y=\frac{2}{m+1}\,.
\end{equation}
\ch{Furthermore, it can be shown that the representation $(1,3)$ satisfies the fusion algebra}
\begin{align}
   \ch{ (1,3)\times (1,3) = (1,1)+(1,3)+(1,5)\,.}
\end{align}
\ch{Since the representation $(1,5)$ gives rise to an irrelevant bulk field (with scaling dimension $\Delta_{(1,5)}=8+\mathcal{O}(y)$), this implies that 
the nearly marginal bulk field $\phi_{(1,3)}$ satisfies a fusion rule of the type \eqref{eq:fusionShort}, which is a necessary condition for the existence of a short RG flow.}
\ch{Additionally, normalizing the two-point function coefficients of all bulk primaries to unity, the three-point function coefficient $\tensor{C}{_{(1,3)(1,3)(1,3)}}(m)$ of three $\phi_{(1,3)}$ bulk fields 
turns out to be \cite{Zamolodchikov:1987ti}}
\begin{equation}
\label{eq:cvvv}
  \tensor{C}{_{(1,3)(1,3)(1,3)}}(m) \equiv \tensor{C}{_{\mathbb{VVV}}} = \frac{4}{\sqrt{3}} \left(1 - \frac{3}{2}y + \mathcal{O}(y^{2})\right)
\end{equation}
in the small $y$ limit (or, equivalently, in the large $m$ limit by using \eqref{eq:ym}). \ch{Since this is non-zero, all conditions for the existence of a nearby fixed point of the RG flow triggered by $\phi_{(1,3)}$ are met.}
\ch{As we have mentioned earlier, in the pure-bulk case, this RG flow was described already by Zamolodchikov in \cite{Zamolodchikov:1987ti}. His analysis showed that flowing from the $m^{\rm th}$ minimal model in the UV, one ends up with the $(m-1)^{\rm th}$ minimal model in the IR. This results in a change in the central charge}
\begin{equation}
    c^{\ast}-c=-\frac{3}{2}y^{3}+\mathcal{O}(y^4)\,,
\end{equation}
\ch{which starts at cubic order in $y$.}

Regarding the open-string background, \ch{we will assume that the worldsheet matter boundary state factorizes into a minimal-model boundary state and some boundary state of the auxiliary CFT.} The most general \ch{elementary} boundary state in a diagonal Virasoro minimal model can be labeled by two integers $\boldsymbol{a}\coloneqq(a_1, a_2)$, which run over the same range as the Kac labels $(r, s)$ of highest-weight representations and are subject to the same identifications. \ch{As argued by Cardy \cite{Cardy:1989ir},} these boundary states can be expanded as
\begin{equation}
    \|\boldsymbol{a}\rangle \! \rangle _m= \sum_{(r,s)}\frac{S_{(a_1,a_2)}^{(r,s)}}{\sqrt{S_{(1,1)}^{(r,s)}}} |r,s \rangle \! \rangle \,,\label{eq:Bexp}
\end{equation}
\ch{where the Ishibashi states $|r, s \rangle \! \rangle$ \cite{Ishibashi:1988kg} are normalized as $\lim_{\tau\to\infty}\langle \!\langle r, s |e^{-2\pi \tau(L_0+\bar{L}_0-\Delta_{(r,s)})} | r', s' \rangle \!\rangle = \delta_{r,r'}\delta_{s,s'}$,} and $S_{(a_1,a_2)}^{(r,s)}$ are the components of the modular $S$-matrix 
\begin{equation}
    S_{(a_1,a_2)}^{(r,s)}(m)=(-1)^{1+a_1 s+a_2 r} \sqrt{\frac{8}{m(m+1)}}\sin\left(\frac{m+1}{m}\pi a_1 r\right)\sin\left(\frac{m}{m+1}\pi a_2 s\right)\,.
\end{equation}
\ch{To diagnose the effect of a short bulk RG flow on a boundary state, it will be crucial to investigate the structure of the bulk-boundary OPE of the perturbing bulk operator. In particular, for the $\phi_{(1,3)}$ bulk field in the presence of a general Cardy boundary state $\boldsymbol{a}$, we can write}
\begin{equation}
  \phi_{(1,3)}(x+is,x-is)\stackrel{s\to 0}{\sim}\frac{1}{(2s)^{2(1-\frac{2}{m+1})}} \frac{B^{(\boldsymbol{a})}_{(1,3)(1,1)}(m)}{g^{(\boldsymbol{a})}(m)}+\frac{1}{(2s)^{1-\frac{2}{m+1}}}\frac{B^{(\boldsymbol{a})}_{(1,3)(1,3)}(m)}{g^{(\boldsymbol{a})}(m)}\,\psi_{(1,3)}(x)+{\rm reg}\,.\label{eq:BBOPEMM}
    \end{equation}
\ch{where $\psi_{(1,3)}\coloneqq \mathbbm{v} $ is the boundary field transforming in the $(1,3)$ representation of Virasoro algebra.}
\ch{Here $g^{(\boldsymbol{a})}$ denotes the $g$-function (boundary entropy) of the boundary state, which can be identified with the coefficient in the expansion \eqref{eq:Bexp} in front of the $(1,1)$ (identity) Ishibashi state. This means that we can write}
\begin{align}
     \ch{g^{(\boldsymbol{a})}(m)\coloneqq \langle 0 \|\boldsymbol{a}\rangle\!\rangle_{m}=\frac{S_{(a_1,a_2)}^{(1,1)}(m)}{\sqrt{S_{(1,1)}^{(1,1)}(m)}}\,.}\label{eq:gMM}
\end{align}
\ch{Second, the coefficient $B^{(\boldsymbol{a})}_{(1,3)(1,1)}$ of the identity channel in the bulk-boundary OPE \eqref{eq:BBOPEMM} can be equated with the coefficient of the $(1,3)$ Ishibashi states in \eqref{eq:Bexp}, meaning that in terms of the modular $S$-matrix, we can write}
\begin{align}
    \ch{\frac{B^{(\boldsymbol{a})}_{(1,3)(1,1)}(m)}{g^{(\boldsymbol{a})}(m)}\coloneqq \frac{\langle \phi_{(1,3)}\|\boldsymbol{a}\rangle\!\rangle_m}{g^{(\boldsymbol{a})}(m)}=\sqrt{\frac{S_{(1,1)}^{(1,1)}(m)}{S_{(1,1)}^{(1,3)}(m)}}\frac{S_{(a_1,a_2)}^{(1,3)}(m)}{S_{(a_1,a_2)}^{(1,1)}(m)}\,.}\label{eq:BvMM}    
\end{align}
\ch{Finally, the OPE coefficient $B^{(\boldsymbol{a})}_{(1,3)(1,3)}(m)$ in front of $\psi_{(1,3)}$ can generally be computed using the 2-point function bootstrap on the upper-half plane. For arbitrary $m$, this was achieved in \cite{Runkel:1998he}, where all structure constants of the A-series of Virasoro minimal models were expressed in terms of the fusing matrix $\mathsf{F}$ and the modular $S$-matrix. In appendix \ref{app:A}, we derive the corresponding crossing equation in the strict $m\to\infty$ limit (see \eqref{eq:BBcrossing}), which suffices for our needs in this paper. 
In particular, note that for the boundary states of the type $\boldsymbol{a}=(1,a_2)$, the representation $(1,3)$ is actually not present in the spectrum of boundary fields, as can be explicitly checked using the Cardy condition. In such cases, the coefficient $B^{(\boldsymbol{a})}_{(1,3)(1,3)}(m)$ identically vanishes to all orders in $y$.

More generally, we will be interested in such boundary states $\boldsymbol{a}$ for which the ratio ${B^{(\boldsymbol{a})}_{(1,3)(1,3)}}/{g^{(\boldsymbol{a})}}$ vanishes at the leading order $\mathcal{O}(y^0)$. Recalling our discussion in section \ref{sec:3}, this will guarantee perturbativity of the boundary deformation induced by the bulk RG flow triggered by $\phi_{(1,3)}$. As analyzed in more detail in appendix \ref{app:A}, setting $B^{(\boldsymbol{a})}_{(1,3)(1,3)}$ to zero in the crossing relation \eqref{eq:BBcrossing} yields two distinct solutions for the coefficient $B^{(\boldsymbol{a})}_{(1,3)(1,1)}$ which, in turn, can be associated with two distinct classes of boundary states for which the bulk-induced boundary deformation remains perturbative. These classes will be described in more detail below.
}

\subsection{Short RG flows in Virasoro minimal models with boundaries}\label{sec:5.2}
In this subsection, we will briefly review the \ch{exhaustive} analysis of \cite{Fredenhagen:2009tn}, where the authors study the RG flows in the Virasoro minimal models with boundaries, which are induced by the nearly-marginal bulk field $\phi_{(1,3)}$, as well as by the nearly-marginal boundary field $\psi_{(1,3)}$ in the limit of large $m$. 
\ch{One of the main results of their work is a diagram which we reproduce in figure \ref{fig:RGflow} and which summarizes the possible RG flows in this setup. In particular, the upper horizontal line in this diagram refers to the minimal model $m$, while the lower horizontal line refers to the minimal model $m-1$. This means that the horizontal arrows indicate pure-boundary RG flows while the vertical ones denote RG flows induced by the bulk field $\phi_{(1,3)}$.
We can notice that the RG flow chain naturally organizes itself horizontally so that the value of $a_2$ increases from right to left. On the other hand, by changing the value of $a_1$, one would generally produce a separate chain.}

\begin{figure}[!htpb]
    \centering
    \includegraphics[width=1\textwidth]{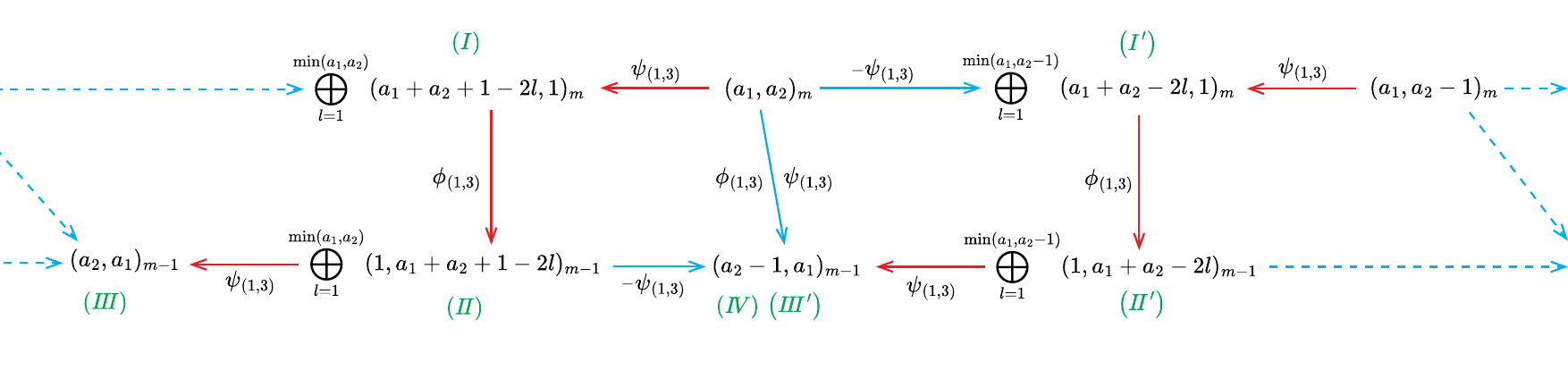}
    \caption{\ch{The chain of RG flows in Virasoro minimal models with boundaries. 
    The upper horizontal line consists of pure-boundary RG flows in the minimal model \texorpdfstring{$m$}{TEXT}, while the lower line contains pure-boundary RG flows in the minimal model \texorpdfstring{$m-1$}{TEXT}. 
    Vertical lines indicate pure bulk or bulk-boundary RG flows which provide the transition from the minimal model \texorpdfstring{$m$}{TEXT} to the minimal model \texorpdfstring{$m-1$}{TEXT}. 
    In particular, here we display the piece of the chain which is generated by RG flows starting from the Cardy boundary state $(a_1,a_2)$ where both $a_1$ and $a_2$ are kept fixed as we take the limit $m\to\infty$. Red lines denote short RG flows, blue lines indicate non-perturbative RG flows, and green Roman numerals mark specific RG fixed points referred to in the text.}
    }
    \label{fig:RGflow}
\end{figure}

\ch{In summary, starting from a boundary state with labels $(a_1,a_2)$ in minimal model $m$, one can identify essentially two distinct sequences of deformations. 

The first one connects the initial boundary state $(a_1,a_2)$ with the fixed points $(I)$, $(I\!I)$ and $(I\!I\!I)$ (see figure \ref{fig:RGflow}). This proceeds 1.\ via a pure-boundary RG flow triggered by $\psi_{(1,3)}$ in the minimal model $m$ which takes us from $(a_1,a_2)$ to the fixed point $(I)$, then 2.\ via a pure-bulk flow induced by $\phi_{(1,3)}$ which is interpolating between the fixed points $(I)$ and $(I\!I)$ and then finally, 3.\ via a pure-boundary flow, again triggered by $\psi_{(1,3)}$ but this time in the minimal model $m-1$, mapping between the fixed points $(I\!I)$ and $(I\!I\!I)$. Identifying the fixed points with specific boundary states in the two respective minimal models, we can write this sequence of deformations as}
\begin{equation}
\label{eq:RGflow1}
    (a_1, a_2)_m\to\bigoplus_{l=1}^{{\rm min}(a_1,a_2)} (a_1+a_2+1-2l,1)_m\to \bigoplus_{l=1}^{{\rm min}(a_1,a_2)} (1,a_1+a_2+1-2l)_{m-1}\to(a_2, a_1)_{m-1}\,. 
\end{equation}
\ch{Crucially, for values $a_1$ and $a_2$ which are kept fixed as we take the limit $m\to \infty$, this sequence of deformations is \emph{perturbative}. Correspondingly, in figure \ref{fig:RGflow} it is highlighted in red color. In more detail, to explicitly confirm perturbativity of the sequence \eqref{eq:RGflow1}, we can use \eqref{eq:gMM} to compute the corresponding change in the $g$-function. This gives}
\begin{align}
\label{eq:gstrag}
    \frac{g^{(a_2,a_1)}(m-1)}{g^{(a_1,a_2)}(m)}=\sqrt{\frac{S_{(1,1)}^{(1,1)}(m)}{S_{(1,1)}^{(1,1)}(m-1)}}\frac{S_{(a_2,a_1)}^{(1,1)}(m-1)}{S_{(a_1,a_2)}^{(1,1)}(m)}=1+\frac{3y}{4}+\frac{21y^2}{32}+\mathcal{O}(y^3)\,,
\end{align}
\ch{which goes to $1$ as $y$ approaches $0$, consistently with our statement. Also notice that since, we are interested in computing the variation of the $g$-function only up to second order in $y$, our analysis will not be sensitive to those steps in the sequence \eqref{eq:RGflow1} which consist of pure-boundary flows, namely steps 1 and 3, which gives contribution at $\mathcal{O}(y^3)$. From the point of view of constructing the corresponding SFT solutions, these two steps would be reflected by turning on a perturbative open-string solution corresponding to a short boundary RG flow both at the level of the initial perturbative background around the boundary state $(a_1,a_2)_m$, as well as at the level of the small fluctuations around the vacuum-shift solution \eqref{eq:OSOL} which describes the fixed point $(I\!I)$.}
\ch{Hence, to the order in $y$ up to which we are working, the on-shell disk action of the open-closed SFT solution computed in section \ref{sec:4} should encode the difference of $g$-functions associated with the deformation}
\begin{equation}
\label{eq:shortRGflow1}
    (a_1, a_2)_m\to (a_2, a_1)_{m-1}\,,
\end{equation}
\ch{which can be viewed as a streamlined version of the step-by-step deformation \eqref{eq:RGflow1}. As we have advertised in the previous section, this class of perturbative deformations is associated with one of the two solutions \eqref{eq:Bsols} for ${B^{(\boldsymbol{a})}_{(1,3)(1,1)}}/{g^{(\boldsymbol{a})}}$ of the crossing equation \eqref{eq:BBcrossing} where we put ${B^{(\boldsymbol{a})}_{(1,3)(1,3)}}$ to be $\mathcal{O}(y)$: indeed, for arbitrary labels $a_1$ and $a_2$, as long as we keep them fixed as we send $m\to \infty$, it follows from \eqref{eq:BvMM} that ${B^{(\boldsymbol{a})}_{(1,3)(1,1)}}/{g^{(\boldsymbol{a})}}=\sqrt{3}+\mathcal{O}(y^2)$.
}

\ch{The second type of deformation in this setup brings us from the boundary state $(a_1,a_2)_m$ to the fixed point $(I\!V)$. 
As we can see on figure \ref{fig:RGflow}, this can proceed either via the fixed points $(I)$ and $(I\!I)$ (from where a pure-boundary flow triggered by $\psi_{(1,3)}$ with the opposite sign must be followed), or, via the fixed points $(I')$ and $(I\!I')$ by first triggering a pure-boundary flow by $-\psi_{(1,3)}$ in the minimal model $m$, then following a pure-bulk flow induced by $\phi_{(1,3)}$ and then finally, riding down yet another pure-boundary flow in the minimal model $m-1$. One can also follow an intermediate path by simultaneously turning a combination of bulk and boundary deformations starting from $(a_1,a_2)_m$. Altogether, this gives us the flow}
\begin{equation}
\label{eq:RGflow2}
    (a_1,a_2)_{m}\to(a_2-1,a_1)_{m-1}\,,
\end{equation}
\ch{which, if we keep the labels $a_1$ and $a_2$ fixed as we take $m\to\infty$, is \emph{non-perturbative}.} Indeed, in this case we have
 \begin{align}
    \frac{g^{(a_2-1,a_1)}(m-1)}{g^{(a_1,a_2)}(m)}=\sqrt{\frac{S_{(1,1)}^{(1,1)}(m)}{S_{(1,1)}^{(1,1)}(m-1)}}\frac{S_{(a_2-1,a_1)}^{(1,1)}(m-1)}{S_{(a_1,a_2)}^{(1,1)}(m)}=1-\frac{1}{a_2}+\frac{3(a_2-1)y}{4a_2}+\mathcal{O}(y^2)\,,
\end{align}
\ch{which does \emph{not} approach $1$ in the small $y$ limit. On the other hand, near the middle of the chain displayed in figure \ref{fig:RGflow}, specifically for $a_2=\frac{m+1}{2}+\alpha$, where we are fixing $\alpha$ as we take $m\to\infty$, perturbativity of the flow \eqref{eq:RGflow2} is achieved, as can be seen by analyzing the $g$-functions}
\begin{align}
\label{eq:gstargmidle}
 \frac{g^{((m-1)/2+\alpha, a_1)}(m-1)}{g^{(a_1, (m+1)/2+\alpha)}(m)}=\sqrt{\frac{S_{(1,1)}^{(1,1)}(m)}{S_{(1,1)}^{(1,1)}(m-1)}}\frac{S_{((m-1)/2+\alpha,a_1)}^{(1,1)}(m-1)}{S_{(a_1,(m+1)/2+\alpha)}^{(1,1)}(m)}=1-\frac{y}{4}-\frac{3y^2}{32}+\mathcal{O}(y^3)\,.
\end{align}
\ch{Therefore, near the middle of the chain, a second independent \emph{short} RG flow}
\begin{equation}
    \label{eq:shortRGflow2}
     \left(a_1,\tfrac{m+1}{2}+\alpha\right)_{m}\to\left(\tfrac{m-1}{2}+\alpha,a_1\right)_{m-1}
\end{equation}
\ch{becomes available. It can be readily checked using \eqref{eq:BvMM} that for such values of the labels $a_1$ and $a_2$, the bulk-boundary OPE coefficient of the bulk field $\phi_{(1,3)}$ in front of the boundary identity channel becomes ${B^{(\boldsymbol{a})}_{(1,3)(1,1)}}/{g^{(\boldsymbol{a})}}=-\frac{1}{\sqrt{3}}+\mathcal{O}(y^2)$, which can be identified with the second solution of the crossing equation \eqref{eq:BBcrossing}.
}

\subsection{SFT analysis}\label{sec:5.3}

\ch{Our final goal in this paper will be to confirm the two independent short RG flows \eqref{eq:shortRGflow1} and \eqref{eq:shortRGflow2} up to second order in $y$ by computing the change in the $g$-functions using the SFT framework which we described in great detail in sections \ref{sec:3} and \ref{sec:4}. In conformal perturbation theory, such a calculation would require going to the two-loop level. On the other hand, in our SFT scheme, one only needs to compute manifestly finite one-dimensional integrals, as it is apparent from the r.h.s.\ of the final formula \eqref{eq:finalformula} for the on-shell disk action of the SFT solution describing short bulk RG flows in the presence of conformal boundaries. }

\ch{Also recall that the quantity which the on-shell SFT disk action computes is really the difference in the worldsheet disk partition functions between the corresponding two string backgrounds. As such, it is generally sensitive not only to a change in the boundary-state $g$-function, but also to a possible change in the value of the string coupling constant $g_\mathrm{s}$. This is clearly manifested on the l.h.s.\ of \eqref{eq:finalformula}. Thus, we need to devise a method of separating these two pieces of information. As in the case of the Narain deformations of compact free bosons which we discussed in section \ref{sec:2}, our strategy will be to show that for both of the aforementioned classes \eqref{eq:shortRGflow1} and \eqref{eq:shortRGflow2} of short RG flows in Virasoro minimal models, the SFT on-shell disk action \eqref{eq:finalformula} can indeed be identified with the ratio $\frac{g^\ast}{g}$ of the $g$-functions corresponding to the endpoints of the flows \emph{times} a universal factor, which is the same for both classes of flows (thus independent of the open-string solution $\Psi^\ast(y)$ and only sensitive to the closed-string solution $\Phi^\ast(y)$). This universal factor will then be identified with the ratio $(\frac{g^\ast_\mathrm{s}}{g_\mathrm{s}})^{-1}$. Such a strategy is equivalent to \emph{assuming} that the change in the $g$-function is known for one of the flows and using this knowledge to fix the change in the string coupling constant. Then the SFT on-shell disk action \emph{predicts} the change in the $g$-function for the second flow and thus serves as an indepedent check of the (conjectured) RG flow.
}


\ch{Let us start by quantifying various ingredients needed for evaluation of the on-shell disk action, as given by the final result \eqref{eq:finalformula} of section \ref{sec:4}. Recall that the main non-trivial ingredient entering the subleading order of this calculation was the finite part ${\cal A}^{\rm f.p.}_{TT}$ of the two-point disk amplitude of two slightly off-shell tachyons, introduced in \eqref{eq:ATTfp}.
Given the analysis of appendix \ref{app:A}, we can now substitute the explicit expression for the regulated  correlator \ch{\eqref{eq:2pointREG}} of two bulk fields $\phi_{(1,3)}$ on the UHP, obtaining}
 \begin{equation}
   {\cal A}^{\rm f.p.}_{TT}=-1- \frac{\tensor{B}{_{\mathbb{V}\mathbbm{1}}}}{g}\,\tensor{C}{_{\mathbb{VVV}}}\left(1+2\log 2\right).
 \end{equation}
\ch{At the same time, we recall the result ${\cal A}^{\rm f.p.}_{TTTT}=-8$ of \cite{Scheinpflug:2023lfn} for the finite part of the sphere amplitude of four slightly off-shell tachyons. Finally, we also remember the $y$-expansion \eqref{eq:cvvv} of the three-point structure constant $C_{(1,3)(1,3)(1,3)}$, as well as the the observation that in the case of both flows \eqref{eq:shortRGflow1} and \eqref{eq:shortRGflow2}, the $m\to \infty$ limit of the bulk-boundary coefficient ${B^{(\boldsymbol{a})}_{(1,3)(1,1)}}/{g^{(\boldsymbol{a})}}$ only starts receiving finite $m$ corrections at the order $\mathcal{O}(y^2)$.
 }
 
\ch{In view of these results,} the general formula  \eqref{eq:finalformula} becomes
\begin{align}
\label{eq:formulaforgvariation}
 \ch{   \frac{g^\ast}{g}\bigg(\frac{g_\mathrm{s}^\ast}{g_\mathrm{s}}\bigg)^{-1}=
  \left(1+\frac{\sqrt{3}}{4}\frac{\tensor{B}{_{\mathbb{V}\mathbbm{1}}}}{g}\,y
  +\frac{3}{8}\frac{\tensor{B}{_{\mathbb{V}\mathbbm{1}}}}{g} \left(\frac{1}{\sqrt{3}}+\frac{1}{4}\frac{\tensor{B}{_{\mathbb{V}\mathbbm{1}}}}{g}\right)\,y^2 +\mathcal{O}(y^3)\right)\left( 1+ \frac{3y^2}{32}+\mathcal{O}(y^3)\right)^{-1}\!,}
\end{align}
\ch{where, at this point, the ratio ${\tensor{B}{_{\mathbb{V}\mathbbm{1}}}}/{g}$ can be replaced with its strict $m\to \infty$ limit. Also, we have rearranged the expression so that the $\mathcal{O}(y^2)$ term which does not depend on the boundary conditions is explicitly factored away from the rest, in a hope that we will be able to attribute this universal factor to the ratio $(\frac{g_\mathrm{s}^\ast}{g_\mathrm{s}})^{-1}$ of the old and the new string coupling constant. We will now provide an unambiguous check for this expectation by showing that the first factor on the r.h.s.\ of \eqref{eq:formulaforgvariation} precisely reproduces the ratio $\frac{g^\ast}{g}$ of the $g$-functions of the endpoints of the conjectured flows \eqref{eq:shortRGflow1} and \eqref{eq:shortRGflow2}.
}

\ch{Considering first the flow \eqref{eq:shortRGflow1}, which starts from the boundary state $(a_1,a_2)_m$ with both $a_1,a_2$ fixed as we take $m\to\infty$, we can substitute}
\begin{align}
 \frac{\tensor{B}{_{\mathbb{V}\mathbbm{1}}}}{g}\to\frac{B^{(a_1,1)}_{(1,3)(1,1)}(m)}{g^{(a_1,1)}(m)}=\sqrt{\frac{S_{(1,1)}^{(1,1)}(m)}{S_{(1,1)}^{(1,3)}(m)}}\frac{S_{(a_1,1)}^{(1,3)}(m)}{S_{(a_1,1)}^{(1,1)}(m)}=\sqrt{3}+\mathcal{O}(y^2)
\end{align}
\ch{to show that the first factor in \eqref{eq:formulaforgvariation} evaluates to}
\begin{equation}
\ch{  1+\frac{3y}{4}+\frac{21y^2}{32}+\mathcal{O}(y^3)\,,}\label{eq:EXP1}
\end{equation}
\ch{which stands in precise agreement with the large-$m$ expansion \eqref{eq:gstrag} of the ratio $\frac{g^{(a_2,a_1)}(m-1)}{g^{(a_1,a_2)}(m)}$ of $g$-functions for the expected endpoints of the flow \eqref{eq:shortRGflow1}. }

\ch{Similarly, we can consider the remaining class of flows \eqref{eq:shortRGflow2}, for which the UV BCFT is given by the boundary state $(a_1,a_2)_m$ with $a_2 = \frac{m+1}{2}+\alpha$ with both $a_1$ and $\alpha$ kept fixed in the limit $m\to \infty$. In this case, substituting}
 \begin{equation}
     \frac{\tensor{B}{_{\mathbb{V}\mathbbm{1}}}}{g}\to\frac{B^{(a_1,(m+1)/2+\alpha)}_{(1,3)(1,1)}(m)}{g^{(a_1,(m+1)/2+\alpha)}(m)}=\sqrt{\frac{S_{(1,1)}^{(1,1)}(m)}{S_{(1,1)}^{(1,3)}(m)}}\frac{S_{(a_1,(m+1)/2+\alpha)}^{(1,3)}(m)}{S_{(a_1,(m+1)/2+\alpha)}^{(1,1)}(m)}=-\frac{1}{\sqrt{3}}+\mathcal{O}(y^2)\,,
 \end{equation}
\ch{into the first factor in \eqref{eq:formulaforgvariation} yields}
\begin{align}
    \ch{1-\frac{y}{4}-\frac{3y^2}{32}+\mathcal{O}(y^3)\,,}\label{eq:EXP2}
\end{align}
\ch{which exactly matches the large-$m$ expansion \eqref{eq:gstargmidle} of the ratio $\frac{g^{((m-1)/2+\alpha, a_1)}(m-1)}{g^{(a_1, (m+1)/2+\alpha)}(m)}$.
}

\ch{The results \eqref{eq:EXP1} and \eqref{eq:EXP2} confirm that the value of the SFT on-shell disk action factorizes 1.~into the ratio $\frac{g^\ast}{g}$ of the $g$-functions of the initial BCFT and the BCFT describing the classical solution and 2.\ into the ratio $(\frac{g_\mathrm{s}^\ast}{g_\mathrm{s}})^{-1}$ of the corresponding string coupling constants.} At the same time, the latter can be isolated as
\begin{equation}
   \frac{g_\mathrm{s}^\ast}{g_\mathrm{s}}=1+\frac{3y^2}{32}+\mathcal{O}(y^3)=1+\frac{1}{8}\left( \frac{2y}{\tensor{C}{_{\mathbb{VVV}}}} \right)^2+\mathcal{O}(y^3)=1+\frac{1}{8}t(y)^{2}+\mathcal{O}(y^3)\,.\label{gs-minimal}
\end{equation}
\ch{Finally, we note that the quantity $t(y)^{2}$ can be identified with the sphere two-point function of the deforming matter CFT operator $t(y)\,\mathbb{V}$. Thus, comparing \eqref{gs-minimal} with \eqref{narain-gs}, we conclude that, at least to the leading order in $y$, the change in the string coupling constant is described by the \emph{same} expansion in terms of the sphere two-point function of the deforming operator as in the case of Narain deformations.}


\section{Discussion}\label{sec:6}
In this paper we have extended  our understanding of how D-branes adapt to a perturbative change in the closed string background, in the context of (bosonic) open-closed SFT. 

As a first main result we have computed exactly at next-to-leading order the on-shell disk action that is generated by a short bulk RG-flow in the initial matter CFT. As already discussed in \cite{cosmo}, this quantity \ch{receives contributions both} from the closed SFT solution describing the change of \ch{the} closed string background, \ch{as well as from} the corresponding open-string vacuum-shift solution.  In quite \ch{a} full generality, this observable is given by eq.\ \eqref{main1}. This formula is valid under the assumption that the barely relevant bulk field, which is deforming the closed string background, has a sub-leading bulk-boundary OPE with the corresponding barely-relevant boundary field. \ch{This ensures that} the \ch{main} driving force \ch{triggering} the open-string background deformation \ch{is provided by} the boundary-identity contribution to the bulk-boundary OPE of the relevant deforming field in the bulk. 

\ch{Broadening our perspective, the analysis presented in this paper} has confirmed and reinforced the observation already made  in \cite{cosmo} that, although the bulk deformation happens exclusively at the \ch{level of matter CFT}, the solution nevertheless changes the string coupling constant, even if the ghost dilaton is not switched on  (at least up to the perturbative order where our analysis is performed). We have now evidence that both in the case of exactly marginal deformations and of short RG flows triggered by a deforming matter field $T$, there is always a $\mathcal{O}(T^2)$ change in the string coupling constant, \ch{which is universally} proportional to the sphere two-point function of the deforming bulk field as shown in eq.~\eqref{main2}. 

There are several questions which naturally arise.
\begin{itemize}
\item The change in the string coupling constant is a closed-string effect. \ch{While we} have used D-branes as probes to detect \ch{this change}, it should be possible to capture it \ch{within the framework of the} closed SFT alone. Any alternative method to access this quantity in purely closed SFT would thus be welcome. \ch{A} standard way for accessing it would be the computation of a 3-point amplitude in the deformed background which would however effectively require a 5-point amplitude (two deforming fields and 3 external legs) in the original theory. Perhaps simpler and more accessible quantities exist.
\item The ghost-dilaton $D$ and related excitations are expected to play an important role. To start with, one can \ch{envisage correcting the observed change in the string coupling constant induced by a matter deformation} by \ch{adding} \ch{to the solution} a second order correction proportional to the ghost-dilaton. In addition, \ch{in case of relevant bulk RG-flows}, the \ch{(BPZ-dual) ghost-dilaton field}  $\tilde D$ and its trivializing field $\Theta$ are needed to cancel the third order obstruction to the solution induced by the change in the central charge. It would be interesting to better understand the role of these fundamental ghost fields, which do not seem \ch{to have their counterparts} in the pure \ch{2d} CFT description.
\item The status of the \ch{on-shell closed-string} field theory action in presence of non-trivial ghost excitations \ch{(as the ones mentioned in the previous point)} remains largely unexplored. Ideally, the \ch{on-shell} closed SFT action \ch{of a short RG flow solution} could furnish an independent way of \ch{measuring} a change in the string coupling constant, possibly together with a shift in the matter central charge \cite{Scheinpflug:2023lfn}.\footnote{This is provided that the on-shell action is not identically vanishing \cite{Erler:2022agw} thanks to some subtle \ch{boundary-term} contribution.} It would be exciting to make progress in this direction.
\item At the moment, (open-closed) SFT is the only fully consistent framework where to address these questions. However, recent progress from 
the point of view of sigma-model deformations has been reported in \cite{Ahmadain:2024hgd, Ahmadain:2022eso, Ahmadain:2022tew}. \ch{It} would be interesting to see how some of our results and questions get translated into this other approach.
\end{itemize}
We hope \ch{that} our paper \ch{furnishes} a direction \ch{which may prove useful} for future explorations of these basic aspects of string theory.


\section*{Acknowledgments}
We thank Marco Meineri, Jaroslav Scheinpflug and Martin Schnabl for useful discussions. We are also grateful to the organizers and participants of the Workshop on Matrix Models and String Field Theory held from May 5 to May 16, 2024 in Benasque, who provided a stimulating environment which enabled completion of the bulk of this work. JV also thanks INFN Turin and CEICO, Institute of Physics, Czech Academy of Sciences for their hospitality during the initial and final stages of this work, respectively. The work of CM  and AR  is partially supported by the MIUR PRIN contract 2020KR4KN2 “String Theory as a bridge between Gauge Theories and Quantum Gravity” and by the INFN project ST$\&$FI “String Theory and Fundamental Interactions”. The work of JV is supported by the ERC Starting Grant 853507.

\appendix

\section{2-point function of 
\texorpdfstring{$\phi_{(1,3)}$}{TEXT} on the UHP}\label{app:A}

In this appendix, we will compute the large-$m$ limit of the 2-point correlation function
\begin{align}
    \frac{1}{g^{(\boldsymbol{a})} }\,\big\langle \phi_{(1,3)}(i,\bar{i})\, \phi_{(1,3)}(is,\bar{is})\big\rangle_\mathrm{UHP}^{(\boldsymbol{a})}\,,\label{eq:2point}
\end{align}
where $\phi_{(1,3)}$ is the primary with Kac label $(1,3)$ in the unitary Virasoro minimal model $m=3,4,\ldots$ and the parameter $s$ runs from $0$ to $1$. The conformal boundary conditions along the real line will be specified by a Cardy boundary state with Kac labels $\boldsymbol{a}=(a_1,a_2)$ and boundary entropy $g^{(\boldsymbol{a})} = \langle 1\rangle_\mathrm{UHP}^{(\boldsymbol{a})}$. To evaluate the correlation function \eqref{eq:2point}, our strategy will be to exploit the existence of a null state at level 3 in the degenerate module $(1,3)$ in order to show that \eqref{eq:2point} satisfies a third-order differential equation. Using the known values of bulk and boundary OPE coefficients for the A-series of Virasoro minimal models, we will then fix a particular solution to this equation which has the correct factorization properties in the bulk and boundary channels.

Let us start by noting that one can use the doubling trick to rewrite \eqref{eq:2point} in terms of a chiral 4-point function on the full complex plane
\begin{align}
    \big\langle \phi_{(1,3)}(z_1)\,\phi_{(1,3)}(z_2)\,\phi_{(1,3)}(z_3)\,\phi_{(1,3)}(z_4)\big\rangle_\mathbb{C} = (z_{12}z_{34})^{-2h} \,\mathcal{G}({\eta})\,,\label{eq:chiral4pt}
\end{align}
subject to assigning $z_1 = i$, $z_2 = -i$, $z_3=is$, $z_4=-is$ and $h=h_{(1,3)} = \frac{m-1}{m+1}=1-y$.
Note that on the right-hand side of \eqref{eq:chiral4pt}, we have made use of global conformal invariance to recast the correlator in terms of a function 
$\mathcal{G}(\eta)$ of the cross-ratio
\begin{align}
    \eta = \frac{z_{12}z_{34}}{z_{13}z_{24}} = -\frac{4s}{(1-s)^2}\,.
\end{align}
We will see below that sometimes, it may be more convenient to work with a related function $\tilde{\mathcal{G}}(\tilde{\eta})\equiv \mathcal{G}(\eta(\tilde{\eta}))$ which is expressed in terms of the redefined parameter
\begin{align}
    \tilde{\eta}(\eta) = \frac{\eta^2}{1-\eta}\,.
\end{align}
One can then exploit the existence of the null-state 
\begin{align}
    \bigg[(h+2)L_{-3}-2L_{-1}L_{-2}+\frac{1}{h+1}(L_{-1})^3\bigg]\big| \phi_{(1,3)}\big\rangle
\end{align}
to infer that the function $\tilde{\mathcal{G}}(\tilde{\eta})$ satisfies the third-order differential equation
\begin{align}
    0&=\big(\tilde{\eta}^3+4\tilde{\eta}^2\big)\frac{d^3\tilde{\mathcal{G}}}{d\tilde{\eta}^3}+\big[  (4-2 h) \tilde{\eta}^2+(10-8 h) \tilde{\eta}\big]\frac{d^2\tilde{\mathcal{G}}}{d\tilde{\eta}^2}+\nonumber\\
    &\hspace{4cm}+\big[(2 - 5 h  - h^2 )\tilde{\eta}+2 - 7 h + 3 h^2  \big]\frac{d\tilde{\mathcal{G}}}{d\tilde{\eta}}+2 h^2 (1 + h) \tilde{\mathcal{G}}\,.\label{eq:DiffG}
\end{align}
Equation \eqref{eq:DiffG} admits a solution for $\tilde{\mathcal{G}}(\tilde{\eta})$ which, for general $h$, can be explicitly written as a linear combination of hypergeometric functions of the type $_3 F_2\,$ in the variable $-\frac{\tilde{\eta}}{4}$. However, as we are really interested only in the leading contribution to the correlator \eqref{eq:2point} as we take the limit $m\to\infty$, we can set $h=1$ and instead consider solving the simpler differential equation
\begin{align}
    0&=\eta\,(1-\eta)^3\,\frac{d^3\mathcal{G}}{d\eta^3}-2 (1-\eta)^2(1+\eta)\,\frac{d^2\mathcal{G}}{d\eta^2}+2(1-\eta)(1-2\eta)\frac{d\mathcal{G}}{d\eta}+4(2-\eta)\mathcal{G}\label{eq:DiffG1}
\end{align}
for $\mathcal{G}(\eta)$. Doing so, we learn that the 2-point correlator \eqref{eq:2point} takes the form of a rational function in $s$
\begin{align}
      &\frac{1}{g^{(\boldsymbol{a})} }\,\big\langle \phi_{(1,3)}(i,\bar{i})\, \phi_{(1,3)}(is,\bar{is})\big\rangle_\mathrm{UHP}^{(\boldsymbol{a})} =\nonumber\\[1.0mm]
      &\hspace{0.4cm}=\frac{1}{s^2(1-s^2)^4}\bigg[K_1 (s^8+28 s^6-314 s^4+28 s^2+1)+K_2(s^8+28 s^6-826 s^4+28 s^2+1)+\nonumber\\[-1.4mm]
      &\hspace{10.0cm}+K_3(s^7+7 s^5+7 s^3+s)\bigg]\,,\label{eq:UnfixedSol}
\end{align}
where $K_1$, $K_2$ and $K_3$ are some integration constants.
These can be fixed by identifying the states propagating in the bulk channel and the boundary channel of the correlator (taking the limits $s\to 1$ and $s\to 0$, respectively, on the r.h.s.\ of \eqref{eq:UnfixedSol}) and then matching the coefficients in front of poles with the corresponding OPE structure constants. 

First, in the bulk channel, one can expand the r.h.s.\ of \eqref{eq:UnfixedSol} as
\begin{align}
    \frac{-16 K_1-48 K_2+K_3}{(1-s)^4}+\frac{32 K_1+64 K_2+K_3}{2 (1-s)^2}+\mathcal{O}\big[(1-s)^{-1}\big]\,.\label{eq:BulkExp}
\end{align}
Here, the quartic pole is to be associated with the propagation of the bulk identity $\phi_{(1,1)}$ in the  bulk channel, while the quadratic pole arises due to propagation of $\phi_{(1,3)}$.
Comparing \eqref{eq:BulkExp} with the result of taking the bulk OPE of the two bulk operators $\phi_{(1,3)}$ inside the correlator \eqref{eq:2point}, we obtain the constraints 
\begin{subequations}
    \begin{align}
        -1&=16 K_1+48 K_2-K_3\,,\label{eq:bulk1}\\
       \frac{1}{4} C_{(1,3)(1,3)(1,3)}\frac{B_{(1,3)(1,1)}^{(\boldsymbol{a})}}{g^{(\boldsymbol{a})}}&=16 K_1+32 K_2+\frac{K_3}{2}\,.\label{eq:bulk2}
    \end{align}
\end{subequations}
We recall that in the strict limit $m\to \infty$, the bulk 3-point structure constant $C_{(1,3)(1,3)(1,3)}$ is equal to $\frac{4}{\sqrt{3}}$, while the value of the disk one-point function coefficient ${B_{(1,3)(1,1)}^{(\boldsymbol{a})}}$ will generally depend on the particular choice $\boldsymbol{a}$ of the Cardy boundary condition. 

Second, expanding the r.h.s.\ of \eqref{eq:UnfixedSol} in the limit $s\to 0$ (boundary channel), one gets
\begin{align}
    \frac{K_1+K_2}{s^2}+\frac{K_3}{s}+\mathcal{O}(s^0)\,,
\end{align}
where the quadratic and simple poles are due to propagation of the identity and the $(1,3)$ module in the boundary channel, respectively. This prompts us to identify
\begin{subequations}
    \begin{align}
        \frac{1}{16}\bigg[\frac{B_{(1,3)(1,1)}^{(\boldsymbol{a})}}{g^{(\boldsymbol{a})}}\bigg]^2 &=K_1+K_2\,,\label{eq:boundary1}\\
        \frac{1}{4}\bigg[\frac{B_{(1,3)(1,3)}^{(\boldsymbol{a})}}{g^{(\boldsymbol{a})}}\bigg]^2&= K_3\,,\label{eq:boundary2}
    \end{align}
\end{subequations}
where ${B_{(1,3)(1,3)}^{(\boldsymbol{a})}}$ denotes the coefficient which enters the two-point function between a $(1,3)$ field in the bulk and a $(1,3)$ field on the boundary.

By solving the constraints \eqref{eq:bulk1}, \eqref{eq:bulk2} and \eqref{eq:boundary1}, one can fix the integration constants $K_1$, $K_2$ and $K_3$ (and hence the correlator \eqref{eq:2point}) in terms of $C_{(1,3)(1,3)(1,3)}$, ${B_{(1,3)(1,1)}^{(\boldsymbol{a})}}$ and $g^{(\boldsymbol{a})}$. The remaining constraint \eqref{eq:boundary2} then provides a non-trivial bulk-boundary crossing relation for the structure constant $B_{(1,3)(1,3)}^{(\boldsymbol{a})}$, namely
\begin{align}
    \bigg[\frac{B_{(1,3)(1,3)}^{(\boldsymbol{a})}}{g^{(\boldsymbol{a})}}\bigg]^2 = 2-2\bigg[\frac{B_{(1,3)(1,1)}^{(\boldsymbol{a})}}{g^{(\boldsymbol{a})}}\bigg]^2+C_{(1,3)(1,3)(1,3)}\,\frac{B_{(1,3)(1,1)}^{(\boldsymbol{a})}}{g^{(\boldsymbol{a})}}\,.\label{eq:BBcrossing}
\end{align}
Validity of this relation can, in principle,  be checked against the known values of the structure constants in minimal models \cite{Runkel:1998he} in the limit $m\to\infty$.

To facilitate applications to the SFT computation of the change in the $g$-function of Cardy boundary conditions upon introducing a bulk perturbation (section \ref{sec:5}), we should keep in mind our assumption \eqref{eq:assumption}, which ensures that the deformation triggered by the bulk field $\phi_{(1,3)}$ on the boundary remains perturbative. From now on, we will therefore fix the Cardy boundary condition $\boldsymbol{a}$ so that the l.h.s.\ of \eqref{eq:BBcrossing} vanishes. This implies two distinct solutions  
\begin{align}
    \frac{B_{(1,3)(1,1)}^{(\boldsymbol{a})}}{g^{(\boldsymbol{a})}} = \left\{
    \begin{array}{l}
           \sqrt{3}+\mathcal{O}(y)\\[3mm]
          -\frac{1}{\sqrt{3}} +\mathcal{O}(y)
    \end{array}
    \right.\label{eq:Bsols}
\end{align}
of the crossing equation \eqref{eq:BBcrossing} for the structure constant ${B_{(1,3)(1,1)}^{(\boldsymbol{a})}}$.

The first solution $\sqrt{3}$ can be realized by taking the Kac labels $(a_1,a_2)$ of the Cardy boundary state \emph{arbitrary but fixed} when taking the limit $m\to \infty$. Correspondingly, one can establish that for such boundary states, the bulk field $\phi_{(1,3)}$ can indeed be used to trigger a sequence of perturbative bulk and boundary deformations. These turn out to relate the initial boundary condition $(a_1,a_2)$ in the minimal model $m$ to the boundary condition $(a_2,a_1)$ in the minimal model $m-1$. 
This is true in particular for the boundary conditions of the type $(a_1,1)$ for which the module $(1,3)$ is absent from the boundary spectrum so that the structure constant $B_{(1,3)(1,3)}^{(\boldsymbol{a})}$ vanishes identically to all orders in $y$. In these cases, there is a direct RG flow taking the boundary condition $(a_1,1)_m$ to $(1,a_1)_{m-1}$. See \cite{Fredenhagen:2009tn}, as well as section \ref{sec:5} for details.

On the other hand, the second solution $-\frac{1}{\sqrt{3}}$ arises for the boundary states with Kac labels $(a_1,a_2)$, where $a_1$ is kept fixed while we put $a_2 = \frac{m+1}{2}+\alpha$ for $m$ odd (and for a fixed integer $\alpha$). This is in precise agreement with the existence of an RG flow triggered by the bulk field $\phi_{(1,3)}$  taking the boundary condition $(a_1, a_2)_m$ directly to the boundary condition $(a_2-1,a_1)_{m-1}$. In the aformentioned regime of the Kac labels $(a_1,a_2)$, such a flow is indeed perturbative.

Focusing our attention on those structure constants $B_{(1,3)(1,1)}^{(\boldsymbol{a})}$ which solve the crossing equation \eqref{eq:BBcrossing} with zero l.h.s.\ (consistent with our assumption \eqref{eq:assumption}), then, in the strict $m\to\infty$ limit,
the desired two-point correlator can finally be fixed as
\begin{align}
     &\frac{1}{g^{(\boldsymbol{a})} }\,\big\langle \phi_{(1,3)}(i,\bar{i})\, \phi_{(1,3)}(is,\bar{is})\big\rangle_\mathrm{UHP}^{(\boldsymbol{a})}=\nonumber\\
     &\hspace{4cm}=C_{(1,3)(1,3)(1,3)}\,\frac{B_{(1,3)(1,1)}^{(\boldsymbol{a})}}{g^{(\boldsymbol{a})}}\frac{s^4+30 s^2+1 }{32 s^2 \left(s^2-1\right)^2}+\frac{ (s^4+14 s^2+1)^2}{16 s^2 (s^2-1)^4}\,.
\end{align}
Furthermore, subtracting the bulk and the boundary channel divergences in a way described in section~\ref{sec:4}, we obtain
\begin{align}
     &\frac{1}{g^{(\boldsymbol{a})} }\,\Big[\big\langle \phi_{(1,3)}(i,\bar{i})\, \phi_{(1,3)}(is,\bar{is})\big\rangle_\mathrm{UHP}^{(\boldsymbol{a})}-\mathrm{div}_\mathrm{c}(s)-\mathrm{div}_\mathrm{o}(s)\Big]=\nonumber\\
     &\hspace{4.8cm}=\frac{1}{(1+s)^4}+C_{(1,3)(1,3)(1,3)}\,\frac{B_{(1,3)(1,1)}^{(\boldsymbol{a})}}{g^{(\boldsymbol{a})}}\frac{ s+3}{4 (1-s) (1+s)^2}\,.\label{eq:2pointREG}
\end{align}
This is the function which is to be integrated against the $c$-ghost measure $4(s^2-1)$ in \eqref{eq:ATTfp} when calculating the contribution $\mathcal{A}_{TT}^\text{f.p.}$ to the on-shell disk action at subleading order in $y$ in section \ref{sec:5}.


\endgroup

\end{document}